\pdfoutput=1

\documentclass[aps,twocolumn,amsmath,amssymb,preprintnumbers,floatfix,longbibliography]{revtex4-2}
\usepackage[utf8]{inputenc}
\usepackage{newtxtext}
\usepackage[upint]{newtxmath}
\usepackage{microtype}
\usepackage{textcomp}
\usepackage{bm}
\usepackage{siunitx}
\usepackage{comment}

\usepackage{enumerate}
\usepackage{amsfonts}
\usepackage{amsmath}
\usepackage{amssymb}
\usepackage{color}
\usepackage{soul}

\usepackage{graphicx}

\usepackage[colorlinks,allcolors=blue]{hyperref}
\usepackage[capitalize]{cleveref}

\definecolor{DarkRed}{rgb}{0.65,0,0}%
\definecolor{Green}{rgb}{0,0.3,0.3}
\definecolor{Purple}{rgb}{0.3,0,0.65}
\definecolor{Red}{rgb}{1,0,0}
\definecolor{Blue}{rgb}{0,0,0.85}
\definecolor{Magenta}{rgb}{1,0,1}

\newcommand{\ve}[1]{\boldsymbol{#1}}

\newcommand{\vech}{\ve{h}} 

\newcommand{\ca}[2][]{c_{#2}^{\vphantom{\dagger}#1}} 
\newcommand{\cc}[2][]{c_{#2}^{{\dagger}#1}}          

\newcommand{\abs}[1]{|#1|}

\newcommand{\vecsigma}{\boldsymbol{\sigma}}

\newcommand{\veck}{\ve{k}}

\newcommand{\vecr}{\ve{r}}

\newcommand{\vecq}{{\ve{q}}}

\newcommand{\vecu}{\ve{u}}

\newcommand{\vecS}{\ve{S}}

\newcommand{\eq}{Eq.}
\newcommand{\etal}{\emph{et al.}}

\newcommand{\be}{\begin{equation}}
\newcommand{\ee}{\end{equation}}

\usepackage{layouts}
\newcommand{\prlsection}[1]{\textit{#1}.\kern0.05em---\kern0.05em\ignorespaces}

\begin{document}
\title{Transient dynamics and quantum phase diagram for the\\square lattice Rashba-Hubbard model at arbitrary hole doping}
\author{Erik Wegner Hodt}
\email[Corresponding author: ]{erik.w.hodt@ntnu.no}
\affiliation{Center for Quantum Spintronics, Department of Physics, Norwegian \\ University of Science and Technology, NO-7491 Trondheim, Norway}
\author{Jabir Ali Ouassou}
\affiliation{Center for Quantum Spintronics, Department of Physics, Norwegian \\ University of Science and Technology, NO-7491 Trondheim, Norway}
\author{Jacob Linder}
\affiliation{Center for Quantum Spintronics, Department of Physics, Norwegian \\ University of Science and Technology, NO-7491 Trondheim, Norway}

\begin{abstract}
Adding a Rashba term to the Hubbard Hamiltonian produces a model which can be used to learn how spin-orbit interactions impact correlated electrons on a lattice. Previous works have studied such a model using a variety of theoretical frameworks, mainly close to half-filling. In this work, we determine the magnetic phase-diagram for the Rashba-Hubbard model for arbitrary hole doping using a sine square deformed lattice mean-field model with an unrestricted ansatz, thus suppressing finite size effects and allowing for inhomogeneous order. We find that the introduction of Rashba spin-orbit coupling significantly alters the ground state properties of the Hubbard model and we observe an increasing complexity of the ground state phase composition for increasing spin-orbit strength. We also introduce a gradual deformed envelope (GDE) technique building on the sine square methodology to facilitate convergence towards ordered and defect-free ground state configurations which is a challenge with the unrestricted ansatz at high interaction strengths. We observe that the use of the GDE technique significantly lowers the free energy of the obtained configurations. Moreover, we consider transient dynamics in the Rashba-Hubbard model by quenching the interaction strength. We find that the quench dynamics within a sine-square methodology allows for the simulation of quasi-open systems by using the zero-energy edge states as a particle reservoir. Interaction quenches at half-filling show a tendency towards quench-induced spatial spin-magnitude inhomogeneity and a non-equilibrium system magnetization lower than equilibrium predictions, possibly related to a build-up of non-local correlations on the lattice. 
\end{abstract}
\maketitle

\section{Introduction}
Atomic spin-orbit coupling is a relativistic effect of central importance in condensed matter physics. From the reference frame of an electron moving in a crystal, the positively charged lattice ions appear to move in the opposite direction. The resulting electric current creates a magnetic field which then couples to the electron spin. This effect is typically large in heavy metals such as Au and Pt. Additional spin-orbit interactions occur in crystals that have no center of inversion, at interfaces between materials, and in thin films. Because of its prevalence in condensed matter systems, such as those described above, spin-orbit interactions play a key role in several research fields  \cite{manchon_natmat_15, zhai_rpp_15, amundsen_rmp_22}. This ranges from topics pertinent to fundamental physics, such as the emergence of Dirac, Weyl, and Majorana quasiparticles in topological matter, to more practically oriented topics, such as enabling information transfer and detection of spin in solid state devices. 

The magnetic properties of the Hubbard model has been a topic of interest since its inception, and remains to some extent disputed, especially in the presence of spin-orbit coupling. Within a mean-field treatment, the initial works of Penn for the 3D square lattice, reproduced by Hirsch \cite{penn, Hirsh} in 2D, established the commonplace Hubbard three-phase diagram with an antiferromagnetic (AF) phase close to half-filling, a ferromagnetic (FM) phase at higher magnitudes of the Hubbard-\textit{U} interaction and a paramagnetic phase for the doped model at lower interaction strengths. A central prediction of these types of diagrams, namely the persistence of the commensurate AF phase when doped away from half-filling was however quickly disputed by a range of papers \cite{Poilblanc, verges, auerbach, avinash} in the late 80's and early 90's, including by Hirsch himself, finding no tendency towards AF ordering beyond half-filling using a Monte-Carlo technique, indicating the presence of the doped AF phase to be an artifact of the mean-field method. The prediction of phase separation in the model, initially by Vischer \cite{visscher} contributed towards an apparent reputation of ineptitude regarding the ability of mean-field techniques to accurately reflect the model properties. More recent works on magnetism in the Hubbard model \cite{chubukov, Langmann2007, Igoshev2010} has also called attention to the role of negative electron compressibility as an indicator of phase separation and instability of the model, typically close to half-filling and with homogeneous mean-field ansätze, establishing the importance of inhomogeneity in the ground state and the need for caution when using mean-field theories. 

The use of sine-square deformed (SSD) envelope-based techniques on finite atomic lattices was made relevant by C. Hotta and N. Shibata \cite{Hotta2012} in 2012 and has become a valuable tool in the investigation of magnetic properties in many-body systems. While the goal is often to map the properties of systems in the thermodynamic limit, numerical restrictions often require compromises to be made. Periodic boundary conditions are typically used to emulate large structures by imposing translational invariance on the system. This prevents some of the finite size effects that arise in an alternative approach, open boundary conditions. However, it comes at the cost of requiring the size of the unit cell, the periodically repating entitty on the lattice, to be explicitly chosen \textit{a priori}. This introduces bias in the calculations which may obscure the actual model ground state. Open boundary conditions typically entail that the lattice edges are modelled as ``hard walls" through which no particles can propagate. The edge sites are thus coupled only to the sites in the lattice interior. By studying finite size systems with open boundary conditions, we no longer explicitly require the system properties to abide by a fixed lattice periodicity, but at the cost of finite size effects and frustrations introduced by the open boundaries. This is where the SSD technique comes into play. The use of SSD envelopes on finite size lattices allows us to mimick the thermodynamic limit by screening out finite size effects caused by the lattice edges while at the same time imposing no restrictions on the system ordering, be it on the spin or charge distribution.

The majority of previous works on the Hubbard model with Rashba-type spin-orbit coupling has been restricted to the case of half-filling \cite{Sun2017, Minar, Zhang2015, Kawano2023}. Works on the doped Rashba-Hubbard model has been largely absent until the last year. Recently, the magnetic phase diagram of the doped Rashba-Hubbard model was reported using a restricted mean-field methodology by Kennedy \textit{et al.} \cite{Kennedy2022} while Beyer \etal\ \cite{beyer_et} discussed magnetic and superconducting properties of the doped Rashba-Hubbard model using a functional renormalization group study. The magnetic phase diagram of the Rashba-Hubbard model is however still not properly established and even at half-filling, there is some dispute as to the method-dependence of previously found results, as for instance discussed by Kawano \etal\ \cite{Kawano2023}.   

In addition to determining the ground state magnetic ordering in the Rashba-Hubbard model, we determine its response to quantum quenches in the electron interaction strength. Quenching refers to a rapid change in one of the parameters of the Hamiltonian which triggers a dynamical evolution of the system from its equilibrium state to a non-equilibrium excited state \cite{quench_review}. 
Key questions of interest in quenching are related to whether the quenching results in a stationary state and what the time-scale and microscopic origin is of thermalization in quenched systems. 

Such non-equilibrium states can be studied experimentally using, for instance, angle-resolved photoemission spectroscopy (ARPES) and its time resolved version (TR-ARPES). ARPES measurements provide information about the Green function of the system, which in turn reveals the band structure of the system, such as the presence of gaps. A prominent example of the interesting physics that arises out of quantum quenches is excitations of high-temperature superconductors. Experiments have observed  \cite{kaiser_prb_14, hu_natmat_14} metastable superconducting properties in cuprate materials which feature a $d$-wave superconducting order parameter at much higher temperatures than superconductivity could persist under equilibrium conditions. 

Quenching can be performed in several different parameters, including magnetic field and interaction strength  \cite{mitra_arcmp_18}. In this regard, cold-atom systems on tunable optical lattices are useful with regard to experimental tunability since Feschbach resonances can be used to tune interaction strengths whereas the very geometry of the lattice itself can in principle also be quenched. Up until now, quenching in the Rashba-Hubbard model has not been studied to the best of our knowledge.

In this work, we first consider the magnetic ground state properties of the Rashba-Hubbard model on a 2D square lattice with the recently developed sine-square deformed mean-field theory to suppress the effects of open boundary conditions on the ground state configurations. 
We also consider the effect of doping on the magnetic properties of the model and utilize the weak-coupling random-phase approximation as a framework to elucidate the driving mechanism behind the formation of magnetic phases. Secondly, we present results for the behavior of the magnetic configurations following a quench in the on-site interaction strength.

\section{Theory}
\subsection{Rashba-Hubbard model and mean-field theory}
The starting point for this paper is the grand-canonical, spin-$\frac{1}{2}$ Rashba-Hubbard model defined on a $N\times N$ square lattice with nearest-neighbor interactions and open boundary conditions, 
\begin{multline}
H=H_{\text{hop}}+H_{U}+H_{G}\\=\sum_{\langle i,j\rangle,\sigma, \sigma'}\big[t_{ij}\sigma_0^{\sigma\sigma'}-i\alpha_R(\sigma_x \delta_y-\sigma_y \delta_x)^{\sigma\sigma'}\big] \cc{i,\sigma}\ca{j,\sigma'} \\+U\sum_{i}\bigg(n_{i,\uparrow} -\frac{1}{2}\bigg)\bigg(n_{i,\downarrow} -\frac{1}{2} \bigg) -\mu\sum_{i}n_{i}
\end{multline}
where the hopping parameter $t_{ij}=-t$ is assumed isotropic and site-independent. The operator $\cc{i,\sigma}(\ca{i,\sigma})$ creates (annihilates) an electron on site \textit{i} with spin projection $\sigma$ and $U>0$ is the on-site repulsive Hubbard interaction strength. $n_{i,\sigma}=\cc{i,\sigma}\ca{i,\sigma}$ counts the number of electrons with spin $\sigma$ at site \textit{i}. Moreover, $\alpha_R$ represents the Rashba spin-orbit coupling strength, $\sigma_0$ is taken to be the identity matrix in spin space, and $\sigma_{i}$ for $i\in\{x,y,z \}$ are the Pauli matrices. The vectors $\boldsymbol\delta=(\vecr_i-\vecr_j)/a$ connect nearest-neighbour sites \textit{i, j} and $\delta_x (\delta_y)$ is the \textit{x} (\textit{y}) component of this vector. The chosen formulation of $H_U$ retains the particle-hole symmetry of the model, fixing half-filling at $\mu = 0$. Below, we use units where the hopping parameter $t$ and lattice constant $a$ are set to unity.

We now apply the identity $n_{i,\uparrow}n_{i,\downarrow}=n_{i}^2 / 4 - (\vecS_i \cdot \vecu_i)^2$, in effect decoupling the charge and spin degrees of freedom,  and introduce the mean charge and spin expectation fields $\langle n_i \rangle$ and $\langle \vecS_i \rangle$. Here, $\vecu_i$ is an arbitrary unit vector.
The charge operator is defined as $n_i=n_{i,\uparrow}+n_{i,\downarrow}$ and counts the number of electrons on a given site. The spin operator $\vecS_i$ is given by $\vecS_i = \frac{1}{2}\sum_{\sigma,\sigma'}\cc{i,\sigma}\vecsigma^{\sigma\sigma'}\ca{i,\sigma'}$ where $\bm{\sigma} = (\sigma_x, \sigma_y, \sigma_z)$ is the Pauli vector. We choose the unit vector $\vecu_i$ to point along the spin expectation value $\langle \vecS_i \rangle$. The interaction term $H_U$ now becomes
\begin{equation}
H_U = U\sum_{i,\sigma}F_i^{\sigma\sigma}\cc{i,\sigma}\ca{i,\sigma} -U\sum_{i,\sigma}G_i^{\sigma\bar\sigma}\cc{i,\sigma}\ca{i,\bar{\sigma}} \label{eqn: interaction term}
\end{equation}   
where the spin $\bar\sigma$ is the opposite of $\sigma$. We can write out the diagonal coefficient $F_i^{\sigma\sigma}$ and off-diagonal coefficient $G_i^{\sigma\bar\sigma}$ as
\begin{align}
F_i^{\sigma\sigma}&=\frac{1}{2}\big(\langle n_i \rangle - 1\big)\sigma_0^{\sigma\sigma} - \langle S_{i,z} \rangle \sigma_{z}^{\sigma\sigma} \\
G_i^{\sigma\bar\sigma} &= \langle S_{i,x}\rangle \sigma_{x}^{\sigma\bar{\sigma}} + \langle S_{i,y}\rangle \sigma_{y}^{\sigma\bar{\sigma}}.
\end{align}
The introduction of the mean-fields introduce operator-free terms in the Hamiltonian given by
\begin{equation}
H_E = \sum_i -\frac{U}{4}\big(\langle n_i \rangle^2 - 1\big) + U\langle \vecS_i \rangle ^2 
\end{equation}
For a given choice of $\alpha_R$, $U$ and $\mu$ (the hopping parameter $t$ will be set to 1 throughout this paper), the mean charge field $\langle n_i \rangle$ and spin field $\langle \vecS_i \rangle$ must be determined self-consistently. Note how the order parameter on each site can adjust freely without any imposed spatial structure with respect to the behavior of the density or magnetic texture, and that the charge and spin degrees of freedom are decoupled, thus allowing for charge and magnetic order to establish independently of each other. The system electron filling level, to be defined later, will be altered implicitly by the chemical potential. In the end, we are  interested in $n_e(\alpha_R, U)$ in order to draw up a phase diagram, where $n_e$ is the average filling level of the system. While some phase diagram calculations in the past has tackled this by fixing the electron density and pinning it throughout the self-consistency calculations by an \textit{a posteriori} fitting of the chemical potential, this assumption of a homogeneous charge field $\langle n_i \rangle$ pinned to some fixed level $n_e$ is problematic.  By imposing a homogeneous electron filling level on the system, one risks ending up with thermodynamically unstable phases, a typical give-away being a negative electron compressibility. 
A good example of a thermodynamically unstable phase is the previously discussed initial three-phase diagrams in the nearest neighbour square lattice Hubbard model and the persistence of the AF phase when the system is doped. A consideration of the electron compressibility in these systems would likely have revealed the instability of the homogeneous AF phase away from half filling, something which today is well established \cite{Igoshev2010,unstableNeel, unstableNeel2}. Using the chemical potential instead of the filling level as the basic variable in calculations improves the credibility of the mean-field result as no \textit{a priori} assumption is made on the electron density of the system, neither on the spatial distribution of charge nor on the average filling level, both of which is determined self-consistently.

The site-dependent charge and spin fields are found self-consistently using an iterative algorithm where the mean-fields are updated after each iteration. We introduce the density matrix 
\begin{align}
\rho^n = \rho(\{\langle n_i \rangle\}^n , \{\langle \vecS_i \rangle\}^n)=\text{e}^{-\beta H(\{\langle n_i \rangle\}^n,\{\langle \vecS_i \rangle\}^n)}/Z
\end{align}
where $Z=\text{Tr}(\text{e}^{-\beta H(\{\langle n_i \rangle\}^n,\{\langle \vecS_i \rangle\}^n)})$.
The density matrix $\rho^n$ is at each iteration \textit{n} a function of the mean-fields $\{\langle n_i \rangle\}^n , \{\langle \vecS_i \rangle\}^n$ at the same iteration. Here, $\beta=1/k_{\text{B}} T$ is the thermodynamic inverse temperature of the system and the set notation indicates that the mean fields are composed of the mean field values at all sites \textit{i}. In each iteration, the new mean-fields are obtained by evaluating the thermal average 
\begin{equation}
\langle A_i \rangle^{n+1} = \text{Tr}[\rho^n A_i]
\end{equation}
where $\langle A_i \rangle$ is either $\langle n_i \rangle$ or $\langle \vecS_i \rangle$. Due to the unrestricted nature of the fields, the charge and spin fields are free to take on a diverse range of configurations depending on initial conditions and model parameters. In some conditions, likely due to the non-rigidity of the energy levels of the mean-field system, the iterative algorithm becomes stuck in oscillations between two different system configurations. To improve convergence and alleviate instabilities in the self-consistency calculations, especially close to half-filling, a mixing factor $\alpha \in (0.0, 1.0]$ is introduced in the above expression. Thus, the $\textit{n}^\text{th}$ iteration introduces an updated mean-field $\langle A_i \rangle^{n+1}$ defined by
\begin{equation}
\langle A_i \rangle^{n+1} = (1-\alpha)\langle A_i \rangle^n + \alpha \text{Tr}[\rho^n A_i]
\end{equation}

Self-consistency calculations can be viewed as a fixed-point iteration $\langle A_i \rangle^{n+1} = \langle A_i \rangle^{n} + \Delta[\langle A_i \rangle^{n}]$, where we have defined a function $\Delta[\langle A_i \rangle^n] \equiv \mathrm{Tr}(\rho^n A_i) - \langle A_i \rangle^{n}$. Note that $\rho^n$ is implicitly a function of all mean fields $\langle A_i \rangle^n$ via the mean-field Hamiltonian. We can interpret $\Delta[\langle A_i \rangle]$ as the flow of the mean-field $\langle A_i \rangle$ towards an attractive fixed point, and convergence is achieved when we reach this point: $\Delta[\langle A_i \rangle] = 0$. However, for $\alpha > 0$, the modified iteration scheme $\langle A_i \rangle^{n+1} = \langle A_i \rangle^{n} + \alpha \Delta[\langle A_i \rangle^{n}]$ still makes $\langle A_i \rangle$ flow towards the same fixed point, and the convergence criterion 
remains ${\Delta[\langle A_i \rangle] = 0}$. In practice, the criterion of convergence will be a small number $\delta$ serving as the numerical threshold for convergence. 
This scheme is known as \emph{simple mixing} in the literature. \cite{dederichs1983a} Adjusting $\alpha$ simply changes the rate of change between numerical iterations, where there is a trade-off between rapid convergence (large $\alpha$) and numerical stability (small $\alpha$). Choosing $\alpha < 1$ reduces the risk of overshooting the real fixed point when updating the mean-fields, which can cause oscillations around the fixed point and thus numerical instability. We found $\alpha = 0.25$ to provide a good trade-off between convergence rate and numerical stability.

In calculations of the magnetic ground state phase diagram, a small finite thermodynamic inverse temperature $\beta=1/k_{\text{B}} T$ with $T=0.01t$ will be used when evaluating Fermi distributions and the thermodynamic potential. In the self-consistency calculations, the initial charge density is set to half-filling across all lattice sites ($\langle n_i \rangle = 1$) while the spin density is randomized both in direction and magnitude across the sites. This is done for each site by drawing $\langle S_i^\mu \rangle$ for $\mu\in\{x,y,z\}$ individually from a uniform distribution  between -$\eta$ and $\eta$ where $\eta$ is $10^{-2}$. 

The free energy of the system after diagonalization is given by
\begin{equation}
F = H_E - \frac{1}{\beta} \sum_{n} \text{ln}(1+\text{e}^{-\beta E_n})
\end{equation}
where $H_E$ contains the constant mean-field terms from the Hamiltonian and $E_n$ are the quasiparticle energy eigenvalues.

\subsection{Sine-square lattice envelope \label{sec: SSD}}
\begin{figure}
    \centering
    \includegraphics{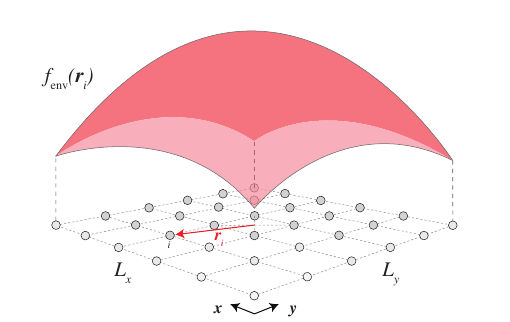}
    \caption{The sine-square deformed envelope for a finite size lattice with $L_x\times L_y$ sites. The envelope modulates the energy scale of the system, screening out the finite size effects associated with the open boundary conditions.}
    \label{fig: envelope}
\end{figure}
In this paper, we consider an $N\times N$ lattice. Calculations of system properties on finite-size lattices with open boundary conditions will always to some extent be affected by the breaking of translational symmetry represented by the edges of the lattice, be it for instance the introduction of Friedel oscillations, affecting the ground state at the lattice center \cite{White2002,Phien2012}. Given the intention of mapping bulk-properties, these edge effects introduce a frustration to the charge and spin configuration which may ultimately disguise the real model ground state due to incompatibilities between the ground state periodicity and the chosen lattice size. While the introduction of periodic boundary conditions alleviates some effects of finite size, the necessary \textit{a priori} selection of lattice periodicity constitutes a significant bias in the search for the appropriate ground state configuration. In the case of incommensurate ordering, we cannot even define an appropriate reduced Brillouin zone due to the irrational relation between the ordering period and the lattice spacing. 
In the Hubbard model, we expect --- based on previous research \cite{Schulz1990, Hiroyuki2016,Halboth2000, Kawano2023} --- a rich ground state behaviour characterized by the presence of incommensurate configurations, i.e. magnetic and charge textures with a periodicity incommensurate with the lattice spacing.

By considering a finite size lattice, energetically modulated by a sine-square envelope, the intention is to include the best of both approaches. By still considering a finite-size system, we do not impose an arbitrary periodicity on the magnetic or charge ordering while simultaneously screen out effects associated with open boundary conditions possibly disguising the appropriate ground state configuration. The introduction of the envelope can, to some extent, be seen as an effective renormalization of the energy scale, causing the edges of the lattice with vanishing energy to serve as a ``particle bath" in analogy with the grand canonical ensemble  \cite{Hotta2012, Hotta2013,Kawano2022}. We might then think of the edge states as a buffer to and from which the system can transfer electrons in order to obtain the optimal ``bulk" filling level in the interior region of the lattice.  We use an envelope function 
\begin{equation}
f_\text{env}(\vecr_i)=\frac{1}{2}\bigg(1+\cos{\frac{\pi\abs{\vecr_i}}{R}} \bigg)
\end{equation}
where $\vecr_i$ is the lattice vector connecting site \textit{i} with the center of the lattice (see Fig. \ref{fig: envelope}). The magnitude \textit{R} is set to the diagonal distance between the lattice center and the corner edge lattice site. 
The sine-square deformed Hamiltonian is then given by 
\begin{widetext}
 \begin{multline}
H=\sum_{\langle i,j\rangle,\sigma,\sigma'}f_\text{env}(\vecr_i, \vecr_j)\big[t_{ij}\sigma_0^{\sigma\sigma'}-i\alpha_R(\sigma_x \delta_y-\sigma_y \delta_x)^{\sigma\sigma'}\big] \cc{i,\sigma}\ca{j,\sigma'} \\+U\sum_{i,\sigma}f_\text{env}(\vecr_i)\bigg[F_i^{\sigma\sigma}\cc{i,\sigma}\ca{i,\sigma} -G_i^{\sigma\bar\sigma}\cc{i,\sigma}\ca{i,\bar{\sigma}}\bigg]-\mu\sum_{i}f_\text{env}(\vecr_i)n_{i}
\end{multline}   
\end{widetext}
where $f_\text{env}(\vecr_i, \vecr_j)=f_\text{env}(\frac{\vecr_i+\vecr_j}{2})$ which we shall abbreviate further to $f_{i,j}\equiv f_\text{env}(\frac{\vecr_i+\vecr_j}{2})$ 
and where the definitions of $F_i^{\sigma\sigma}/G_i^{\sigma\bar\sigma}$ are given with the definition of Eq. (\ref{eqn: interaction term}). 
The operator-free terms are given by
\begin{equation}
H_E = \sum_i f_{i,i} \bigg[-\frac{U}{4}\big(\langle n_i \rangle^2 - 1\big) + U\langle \vecS_i \rangle ^2\bigg] 
\end{equation}

While the presence of open boundary conditions breaks the translational symmetry of the finite size lattice, the system is still translationally symmetric within the cluster. The energetic renormalization of the SSD envelope lifts this symmetry as well. The appearance of reservoir-like states along the edges of the lattice entails that the relevant, ``bulk-like" properties are confined to the interior of the lattice, rendering a traditional unweighted average over all lattice-sites meaningless. Within a SSD framework, we can instead define the average system filling level as \cite{Kawano2022}
\begin{equation}
n_e = \frac{\sum_i f_{i,i}\langle n_i \rangle }{\sum_i f_{i,i}} \label{eqn: avg filling level}
\end{equation}
In order to characterize magnetic ordering, we will calculate the magnetic structure factor. A regular Fourier transform of the magnetic real space texture would yield a non-meaningful result due to the contribution from the edge states. In similar manner as above, we introduce the deformed Fourier transformation \cite{Kawano2022} so that the spin structure factor may be written as 
\begin{equation}
\langle \vecS_{\vecq}\rangle = \frac{\sum_i f_{i,i}\langle \vecS_i \rangle e^{i\vecq\cdot\vecr_i}}{\sum_i f_{i,i}} \label{eqn: mag struc}
\end{equation}
Finally, we would also like to study the charge ordering of the system apart from the average filling level given by \eq{ \eqref{eqn: avg filling level}} as for instance charge-density waves. We may then, in analogy with the magnetic structure factor, calculate a charge structure factor given by 
\begin{equation}
\langle n_\vecq \rangle = \frac{\sum_{i}f_{i,i}\delta n_i \text{ e}^{i\vecq\cdot\vecr_i}}{\sum_{i}f_{i,i}}
\end{equation}
where we have defined the deviation $\delta n_i=\langle n_i \rangle - n_e$, i.e. the local deviation from the average filling level of the system.
\subsection{Random phase approximation and magnetic susceptibility}
In mean-field systems, a common mechanism causing magnetic ordering is nesting of the Fermi surface, the typical example being the $\boldsymbol{Q}=(\pi,\pi)$ nesting vector in the square-shaped Fermi surface of the half-filled Hubbard model giving rise to the Néel antiferromagnet. This is reflected through the magnetic susceptibility which diverges as the system temperature is lowered, causing an instability towards magnetic ordering. Following the derivation in  \cite{Minar, Kawano2023, mahan2000many}, the magnetic susceptibility in RPA when the system breaks spin-rotational symmetry, as in our case, is given by the $3\times3$ RPA susceptibility matrix
\begin{equation}
    \chi_\text{RPA}(\vecq)=\frac{\chi_0(\vecq)}{I_3-2U\chi_0(\vecq)} \label{eqn: RPA}
\end{equation}
where $\chi_0(\vecq)$ is the $3\times3$ bare magnetic susceptibility matrix and $I_3$ is the identity matrix. The bare magnetic susceptibility of the non-interacting system is given by 
\begin{equation}
\begin{split}
    \chi_0^{\mu\nu}(\vecq)=\frac{1}{N}\sum_{\veck, n,m}S_{n,m}^{\mu}(\veck, \veck+\vecq)& S_{m,n}^{\nu}(\veck +\vecq, \veck) \\ &\cdot F_{n,m}(\veck, \veck+\vecq) \label{eqn: susceptibility integrand}
\end{split}
\end{equation}
where $\mu,\nu \in \{x,y,z\}$ denotes the spatial directions and where $F_{n,m}(\veck, \veck+\vecq)$ is the Lindhard function in the zero-frequency limit,
\begin{equation}
    F_{n,m}(\veck, \veck+\vecq)=\frac{f(E_m(\veck)) - f(E_n (\veck + \vecq))}{E_m(\veck)-E_n(\veck+\vecq)+i\eta} \label{eqn: Lindhard}
\end{equation}
and $S^{\mu}(\veck, \veck+\vecq)$ are the Pauli matrices, transformed by the unitary transformation matrices $U_{\boldsymbol{k}}^{\vphantom{\dagger}}$ and $U_{\veck + \vecq}^{\dagger}$ which diagonalize the non-interacting problem:
\begin{equation}
    S^{\mu}(\veck, \veck+\vecq)=\frac{1}{2}U_{\veck}^{\vphantom{\dagger}}\sigma^{\mu}U_{\veck+\vecq}^\dagger
\end{equation}
In the above equations $f$ is the Fermi-Dirac distribution, $N$ denotes the number of modes in $\veck$-space and $\vecq$ denotes the magnetic ordering vector. $\eta>0$ in the Lindhard function is an infinitesimal convergence factor. Being usually scalar for spin-degenerate systems, the susceptibility matrix becomes $3\times 3$ due to the breaking of SU(2) symmetry associated with the Rashba-term. The above results can be derived using the Kubo formula in linear response where the interactions $U$ are treated as a perturbation in the form of an effective magnetic field after a mean-field approximation. The linear response treatment when treating the interactions within mean-field theory corresponds to a random phase approximation, as has been discussed in previous literature.

By diagonalizing the RPA susceptibility matrix given by Eq. (\ref{eqn: RPA}), the eigenvalues of the RPA susceptibility may be written as $\lambda_\text{RPA}^{i}(\vecq)=\lambda_0^{i}/(1-2U\lambda_0^{i}(\vecq))$ for $i=1,2,3$ where $\lambda_0^{i}$ are the eigenvalues of the bare susceptibility matrix in increasing order. The susceptibility matrix was diagonalised numerically. If we start in the non-interacting system and evaluate $\lambda_\text{RPA}^i$ for an initially infinitesimal \textit{U}, continuously ramping up the interaction strength, at some point \textit{U} is large enough to cause the denominator $(1-2U\lambda_0^{3})$ to vanish, where $\lambda_0^{3}$ is the largest eigenvalue of the bare susceptibility, causing $\lambda_\text{RPA}$ to diverge. A divergent eigenvalue necessitates a divergent element in $\chi_\text{RPA}$, causing the spin expectation value
\begin{equation}
    \langle \vecS_{\vecq} \rangle = \chi_{\text{RPA}}(\vecq)\vech_\vecq
\end{equation}
to diverge for the respective ordering vector $\vecq$ where $\bm{h}_\vecq$ is an infinitesimal magnetic field.
An assumption of the above reasoning is that the \textit{U} required to make $(1-2U\lambda_0^{3})$ diverge is small and within the region of validity for RPA. In addition, the assumption that the largest eigenvalue of the bare susceptibility $\lambda_0^3$ is the first to cause a divergence in $\lambda_\text{RPA}$ necessitates that $2U\lambda_0^3 < 1$.

\subsection{Dynamics of observables due to quantum quench}
We will study the effect of a quench at time $t=0$ in either the interaction parameter $U$ or the spin-orbit coupling strength $\alpha_R$. In effect, we will consider a system defined by the Schrödinger picture Hamiltonian 
$H(t) = H_0= H(U,\alpha_R)$ for $t\leq0$ and $H(t)=H_1=H(U+\Delta_U, \alpha_R+\Delta_{\alpha_R})$ for $t>0$.
At $t=0$, before the quench, the system will be in the equilibrium state defined by the density matrix
\begin{equation}
    \rho_0=\text{e}^{-\beta H_0}/Z_{0}, \qquad Z_0=\text{Tr}[\text{e}^{-\beta H_0}]
\end{equation}
Upon the instantaneous change of the system Hamiltonian, the eigenbasis of the Hamiltonian will change, assuming $[H_1, H_0 ]\neq 0$, with a subsequent redefinition of the system ground state. The density matrix $\rho(t)$ will thus become time-dependent and evolve according to the von Neumann equation
\begin{align}
    i\partial_t \rho(t)&=[H(t), \rho(t)]
\end{align}
The solution to this equation is    \begin{align}
    \rho(t) &= U(t)\rho_0 U^\dagger(t), \label{eqn: density matrix time evolution}
\end{align}
where $U(t)$ is a unitary time evolution operator for the density matrix. For a time-independent Hamiltonian, this reduces to simply $U(t)=\text{e}^{-iHt}$, but at this point we make no such assumption. The time-dependent expectation value of the Schrödinger picture operator $A_S$ may then be written as 
\begin{equation}
    \langle {A}(t) \rangle = \text{Tr}\big[\rho(t) {A}_S \big]
\end{equation}
Let us now substitute \cref{eqn: density matrix time evolution} into the above. Using the cyclic property of the trace, and defining the Heisenberg operator ${A}_H(t) \equiv U^\dagger(t) A_S U(t)$, we then obtain the corresponding equation in the Heisenberg picture
\begin{align}
    \langle \hat{A}(t) \rangle &= \text{Tr}\big[\rho_0 {A}_H(t) \big]. \label{eqn: Heisenberg density matrix}
\end{align}
Here, we have rewritten the expectation value in the Heisenberg picture using Eq. (\ref{eqn: density matrix time evolution}) and the cyclic property of the trace. The temporal evolution of the operator ${A}_H(t)$, and thus the average $\langle {A}(t)\rangle$, is now given by the Heisenberg equation 
\begin{equation}
    i \frac{\text{d}}{\text{d}t} \langle {A}(t)\rangle = \Big\langle \big[{A}_H(t), H_H(t) \big] \Big\rangle\label{eqn: Heisenberg equation}
\end{equation}
assuming no explicit time-dependence in the operator $\hat{A}$. In this equation, $H_H(t)$ is the Heisenberg picture Hamiltonian and is related to the Schr{\"o}dinger picture Hamiltonian by the unitary transformation $H_H(t)=U^\dagger(t)H(t)U(t)$ where $H(t)$ is given above. An evaluation of Eq. (\ref{eqn: Heisenberg density matrix}) or Eq. (\ref{eqn: Heisenberg equation}) would thus require an explicit expression for $U$ and $U^\dagger$ which in general depends on a time-integral over $H(t)$.

Instead of evaluating the Heisenberg equation in Eq. (\ref{eqn: Heisenberg equation}) as it stands, we  replace $H_H(t)$ with $H(t)=H_1$ for $t>0$. Taking the change in the system Hamiltonian to be instantaneous, as modelled by a Heaviside step-function, the quench effectively initializes the quenched system in the ground state of the pre-quenched Hamiltonian. The temporal evolution for $t>0$ can thus be thought of as the evolution of an excited state of the quenched Hamiltonian \cite{Calabrese2012, Moeckel2009}, where the time-evolution is governed by the time-independent post-quench Hamiltonian $H_1$, in effect an initial value problem. This quench protocol is reasonable as long as the Hamiltonian changes on a time scale significantly shorter than other relevant time scales in the system. 

In the system discussed in this paper, $\hat{A}$ is either a number operator $\cc{i,\sigma}\ca{i,\sigma}$, spin-conserving hopping operator $\cc{i,\sigma}\ca{i+\delta,\sigma}$ or spin-flipping hopping operator $\cc{i,\sigma}\ca{i+\delta, \bar{\sigma}}$. Evaluating the Heisenberg equation (Eq. (\ref{eqn: Heisenberg equation})) with the post-quench Hamiltonian $H_1$ as discussed above, we obtain three distinct types of dynamical equations for the three types of operators,
\begin{widetext}
\begin{multline}
  i\frac{\text{d}}{\text{d}t}\big\langle \cc{h,\alpha} \ca{h,\alpha} \big\rangle = -t\sum_{\delta} \bigg\{f_{h,h+\delta}\big\langle \cc{h,\alpha}\ca{h+\delta, \alpha} \big\rangle -  f_{h-\delta,h}\big\langle\cc{h-\delta, \alpha}\ca{h,\alpha}\big\rangle \bigg\} \\+ i\alpha_R\sum_{\delta}\bigg\{f_{h,h+\delta} E^{\alpha\bar\alpha} \big\langle \cc{h,\alpha}\ca{h+\delta, \bar\alpha} \big\rangle  - f_{h-\delta,h} E^{\bar\alpha\alpha}\big\langle\cc{h-\delta, \bar\alpha}\ca{h,\alpha}\big\rangle \bigg\} \label{eqn: number equation}
\end{multline}
\begin{multline}
i\frac{\text{d}}{\text{d}t}\big\langle \cc{h,\alpha} \ca{h+\Delta,\alpha} \big\rangle = -t\sum_{\delta}
\bigg\{f_{h+\Delta,h+\Delta+\delta} \big\langle \cc{h,\alpha} \ca{h+\Delta+\delta, \alpha} \big\rangle  -f_{h-\delta,h} \big\langle \cc{h-\delta,\alpha} \ca{h+\Delta, \alpha} \big\rangle\bigg\} \\+ i\alpha_R\sum_{\delta}\bigg\{f_{h+\Delta,h+\Delta+\delta} E^{\alpha\bar\alpha} \big\langle \cc{h,\alpha} \ca{h+\Delta+\delta, \bar\alpha} \big\rangle  - f_{h-\delta,h} E^{\bar\alpha\alpha}\big\langle \cc{h-\delta,\bar\alpha} \ca{h+\Delta, \alpha} \big\rangle \bigg\} \\
+f_{h+\Delta,h+\Delta}\big[F_{h+\Delta}^{\alpha\alpha} \big\langle \cc{h,\alpha}\ca{h+\Delta, \alpha}\big\rangle - G_{h+\Delta}^{\alpha\bar\alpha}\big\langle\cc{h,\alpha}\ca{h+\Delta, \bar\alpha}\big\rangle\big] -f_{h,h}\big[F_h^{\alpha\alpha}\big\langle \cc{h,\alpha}\ca{h+\Delta, \alpha}\big\rangle - G_h^{\bar\alpha\alpha}\big\langle\cc{h,\bar\alpha}\ca{h+\Delta,\alpha}\big\rangle\big] \label{eqn: spin-hop equation}
\end{multline}
\begin{multline}
i\frac{\text{d}}{\text{d}t}\big\langle \cc{h,\alpha} \ca{h+\Delta,\bar\alpha} \big\rangle = -t\sum_{\delta}\bigg\{f_{h+\Delta,h+\Delta+\delta} \big\langle \cc{h,\alpha} \ca{h+\Delta+\delta, \bar\alpha} \big\rangle -f_{h-\delta,h} \big\langle \cc{h-\delta,\alpha} \ca{h+\Delta, \bar\alpha} \big\rangle\bigg\} \\+ i\alpha_R\sum_{\delta}\bigg\{f_{h+\Delta,h+\Delta+\delta} E^{\bar\alpha\alpha} \big\langle \cc{h,\alpha} \ca{h+\Delta+\delta, \alpha} \big\rangle  - f_{h-\delta,h} E^{\alpha\bar\alpha}\big\langle \cc{h-\delta,\bar\alpha} \ca{h+\Delta, \bar\alpha} \big\rangle \bigg\} \\
+f_{h+\Delta,h+\Delta}\big[F_{h+\Delta}^{\bar\alpha\bar\alpha} \big\langle \cc{h,\alpha}\ca{h+\Delta, \bar\alpha}\big\rangle - G_{h+\Delta}^{\bar\alpha\alpha}\big\langle\cc{h,\alpha}\ca{h+\Delta, \alpha}\big\rangle\big] -f_{h,h}\big[F_h^{\alpha\alpha}\big\langle \cc{h,\alpha}\ca{h+\Delta, \bar\alpha}\big\rangle - G_h^{\bar\alpha\alpha}\big\langle\cc{h,\bar\alpha}\ca{h+\Delta,\bar\alpha}\big\rangle\big] \label{eqn: spin-flip equation}
\end{multline}
\end{widetext}
where we have defined the spin-orbit matrix $E=\sigma_x \delta_y - \sigma_y \delta_x$, where $F_i$ and $G_i$ are as defined in \eq{ (\ref{eqn: interaction term})}, where $\delta$ is the nearest-neighbour vector and finally where \textit{h} is a general site index. In Eq. (\ref{eqn: spin-hop equation})-(\ref{eqn: spin-flip equation}), the site denoted by $h+\Delta$ can be both a nearest-neighbour lattice site, but also a general site farther away on the lattice. Note that $\Delta$ could also be the zero-vector, which in Eq. (\ref{eqn: spin-flip equation}) leads to the dynamic equation for the on-site spin-flip operator. 

At any given time $t$, the system configuration is completely characterized by the set of all possible two-operator expectation values ${A}(t) = \big\{\big\langle \cc{i,\alpha}(t)\ca{j, \beta}(t)\big\rangle\big\}$. We denote this set of observables the \textit{statistical state}. Of central importance is the \textit{initial statistical state}, defined before the quench at $t=0$, denoted by \cite{Yukalov2011}
\begin{equation}
    A_0 = \big\{\big\langle \cc{i,\alpha}(0)\ca{j, \beta}(0)\big\rangle\big\}
\end{equation}
This initial statistical state will serve as the initial values in the temporal evolution of the system, the dynamics of each individual average being determined by the appropriate Heisenberg equation in Eq. (\ref{eqn: number equation})-(\ref{eqn: spin-flip equation}). Note that while the original Hamiltonian only include nearest-neighbour hopping and on-site interaction, the time dynamics require the evaluation and temporal evolution of the entire initial statistical state, including next-nearest hopping operators and beyond. This is due to a property of the hopping-operator commutators of the form $\big[\cc{i,\alpha}\ca{i+\delta, \beta}, H\big]$ depending on hopping operators between the nearest neighbours of $i$ and nearest neighbours of $i+\delta$ which again have to be evolved with their own Heisenberg equations, causing the set of Heisenberg equations to be closed only under the finiteness of the lattice itself. Due to this property, the entire initial statistical state will have to be temporally evolved, a set which for an $N\times N$ lattice involves the temporal evolution of $4N^2$ averages coupled at each time-step. When taking into account the hermitian nature of the statistical state, the number of independent averages to evolve reduces to $2N^4+N^2$.

Finally, let us discuss the numerical treatment of the time evolution.
We have mentioned that the system configuration is fully characterized by the {set of} two-operator expectation values $\{c^\dagger_{i\alpha}(t) c_{j\beta}^{\vphantom{\dagger}}(t)\}$.
If we collect these expectation values into a single vector $\bm{A}(t)$, then \cref{eqn: spin-hop equation,eqn: spin-flip equation,eqn: number equation} can be summarized as an equation $i\bm{A}'(t) = \bm{M}(t) \bm{A}(t)$.
Here, $\bm{M}(t)$ is a time-dependent matrix with components given by the envelope $f_{ij}$, hopping $t$, Rashba coefficient $\alpha_R$, as well as the mean-field coefficients $E^{\alpha\bar{\alpha}}, F^{\alpha\bar{\alpha}}_{i}, G^{\alpha\bar{\alpha}}_{i}$ defined previously.
This equation was then solved numerically for a $24 \times 24$ lattice using a 4th-order Runge-Kutta method.
There are however two important considerations to keep in mind.
Firstly, while this looks like a linear differential equation, it is implicitly a nonlinear differential equation.
This is because the coefficients $F$ and $G$ introduced in \cref{eqn: interaction term} are themselves defined in terms of the same mean fields we collected in $\bm{A}(t)$.
This essentially makes our equation of the form $i\bm{A}'(t) = \bm{M}[\bm{A}(t)] \bm{A}(t)$, where in practice evaluate $\bm{M}(t) = \bm{M}[\bm{A}(t)]$ once per time step.
Secondly, we note that $\bm{M}(t)$ is a sparse matrix, which means that the computational effort can be reduced significantly by not constructing it numerically as a dense matrix.
Notably, the matrix $\bm{M}(t)$ has $\mathcal{O}(N^4)$ non-zero elements and $\mathcal{O}(N^8)$ zero elements, making the matrix extremely sparse as $N$ increases.
This sparsity is a result of the Hubbard model only having on-site and nearest-neighbor interactions.
This locality is evident in \cref{eqn: spin-hop equation,eqn: spin-flip equation,eqn: number equation}, where e.g. the time evolution of $\langle c^{\dagger}_{h,\alpha} c^{\vphantom{\dagger}}_{h+\Delta,\alpha} \rangle$ only depends on components like $\langle c^{\dagger}_{h-\delta,\bar{\alpha}} c^{\vphantom{\dagger}}_{h+\Delta,\alpha} \rangle$ that are at most one site ($h \rightarrow h-\delta$) and one spin flip ($\alpha \rightarrow \bar{\alpha}$) away.

\begin{figure*}
\centering
\includegraphics[width=\linewidth]{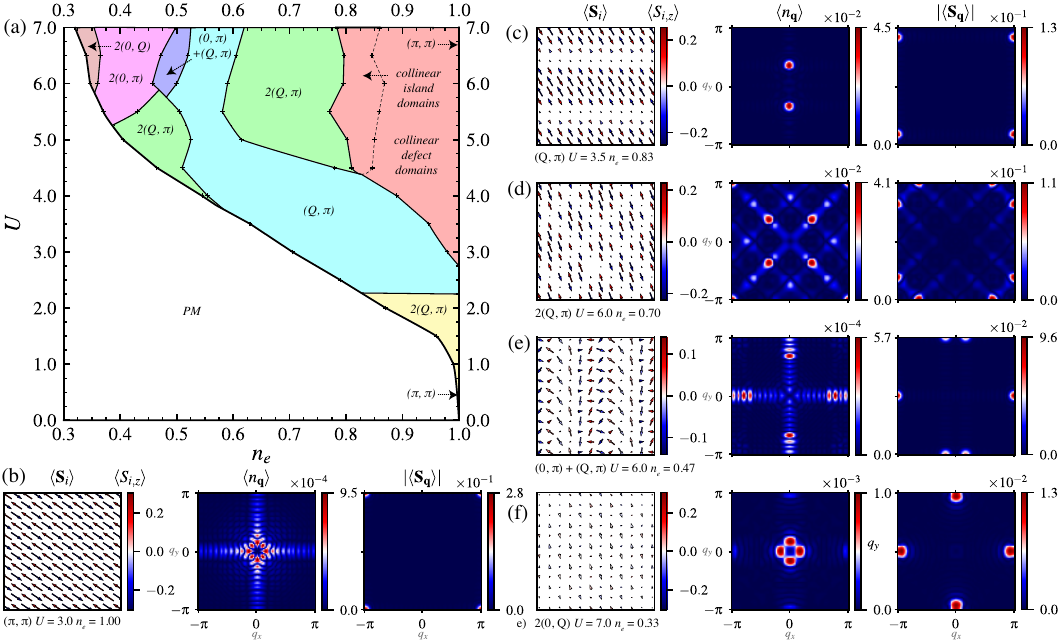}
\caption{Ground state magnetic phase diagram obtained for a $24\times24$ lattice in the absence of spin-orbit coupling ($\alpha_{{R}}=0$) for the SSD model. The phases are classified by the type of ordering vector $\boldsymbol{Q}=(Q_x, Q_y)$ dominating the magnetic structure factor $|\langle \boldsymbol{S}_{\boldsymbol{q}} \rangle|$. An ordering vector component $Q_{x/y}=\pi$ indicates that the component lies on the edge of the first Brillouin zone. If the component takes on the general value $Q_{x/y}=Q$, this indicates a component in the interior of the Brillouin zone, generally incommensurate with the lattice unless specified. The presence of a factor 2 in front of the ordering vector, e.g. $\bm{Q}=2(Q,\pi)$, represents a doubling of the number of maxima in the magnetic structure factor, i.e. describing an increase in ordering symmetry. The collinear island and defect domains are a special region of the phase diagram where the system is characterized by the formation of unordered or semi-ordered regions of charge depletion, bearing the resemblance to lattice defects (see Fig. \ref{fig: defect region}). (b--f)~Examples of the orders shown in the phase diagram in (a), with classification and parameters listed under each plot. 
}
\label{fig: phase diagram alpha=0.0}
\end{figure*}
\subsubsection{Transient dynamics in the SSD-model}
For a system with open boundary conditions, the dynamical equations laid out in the preceding section preserves the particle number. In essence, while the static, self-consistency calculations associated with the magnetic phase diagram involves the coupling to an external particle reservoir through the chemical potential, the temporal evolution of these states happens as a closed system, the number of electrons present in the system being restricted to what the initial statistical state dictates. As such, in the absence of the SSD envelope, a quench in $U$ or $\alpha_R$ can never cause a change to the filling level of the system, only to the charge and magnetic order of the original electron population. This changes upon the introduction of the SSD envelope and the notion of edge states serving as a particle bath. As discussed previously, the presence of electrons along the edges of the lattice with energies approaching zero allows the system to tune the filling level in the ``bulk" interior of the lattice by transferring electrons in and out of this interior region. The effect of this local particle reservoir is that also in the dynamical case are we able to model quench-induced changes to the filling level of the system, by a transfer of electrons to and from the zero-energy edge states as the changes in model parameters alters the energetics of the initial statistical state. This entails that we can, to some extent, model open quantum systems dynamically without taking into account an external particle reservoir explicitly. Note that even in the presence of SSD, the actual particle number is still conserved. The difference is however that we in the presence of SSD draw a distinction between electrons in the interior versus those at the edges, causing the migration of electrons between these two regions to effectively constitute a change in the filling level of the interior region. This places a restriction on the amount of electrons the edge can ``store" or supply to the bulk system and thus the degree to which the edges can act as a reservoir.

\section{Results and discussion}
\subsection{Quantum phase diagram}
\subsubsection{Phase diagram in the absence of Rashba spin-orbit coupling – $\alpha_R = 0.0$}

The phase diagram in the absence of spin-orbit coupling ($\alpha_R=0.00$) was calculated for a $24\times24$ lattice using the SSD Hamiltonian. The phase diagram is shown in Fig. \ref{fig: phase diagram alpha=0.0} and the phases are characterized by the dominant ordering vector $\boldsymbol{Q}$ in their magnetic structure factor $|\langle \boldsymbol{S}_\vecq \rangle|$. The presence of a factor 2 in front of the ordering vector, e.g. $\bm{Q}=2(Q,\pi)$, represents a doubling of the number of maxima in the magnetic structure factor, i.e. describing an increase in ordering symmetry. In the phase diagram, the different phases are also described by a charge structure factor $\langle n_\vecq \rangle$ describing the charge modulation of the phase, relative to the average filling level. We note that due to the definition of the magnetic structure factor (Eq. \ref{eqn: mag struc}), the magnetic ordering vector $\boldsymbol{Q}$ is affected by both the relative orientation between adjacent spins on the lattice (spiral, Néel, stripe etc.), but also by the charge modulation on the lattice, typically giving a resulting spin magnitude modulation. A maximum change in site magnetization or site charge of $\delta=10^{-4}$ between successive iterations was used as a criterion for convergence and the phases were termed paramagnetic (PM) when the average system magnetization of the converged phases fell below $10^{-2}$. More details on the self-consistency calculations are discussed in Sec.~\ref{sec: SF}. 

\begin{figure}
    \centering
    \includegraphics{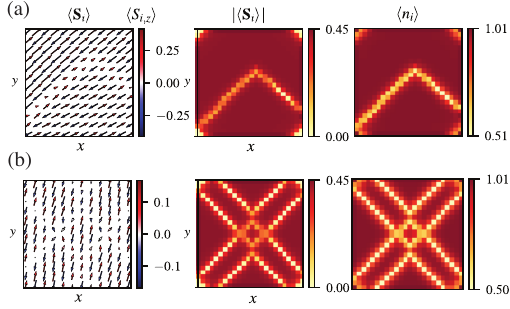}
    \caption{The spin expectation distribution as well as the spin expectation magnitude is shown for the defect region in Fig. (\ref{fig: phase diagram alpha=0.0}). (a) $U=7.0$, $n_e = 0.95$ – Upon reduction of the system filling level, charge defects arise in the lattice, indicated by the inhomogeneoeus charge distribution $\langle n_i \rangle$. The modulation of the spin magnitude $|\langle \boldsymbol{S}_i \rangle|$ is directly linked to these regions of lower charge.  (b) $U=7.0$, $n_e = 0.85$ – When further decreasing the filling level, the charge defects become more ordered, showing a strong tendency to order according to the underlying lattice. 
    }
    \label{fig: defect region}
\end{figure}

The phase diagram displays a variety of charge and spin orders which mainly can be divided into two distinct categories, a region where charge defect formation dominates the system behaviour and thus disrupt the magnetic configurations, and a region with clearly defined phases. The main characteristic of the prior is the formation of charge defects on the lattice with subsequent alterations to the lattice magnetization, shown in more detail in Fig. \ref{fig: defect region}. This phenomenon is prevalent for higher interaction strengths ($U \gg 1$) close to half-filling ($n_e \rightarrow 1$), denoted by the red defect area in the phase diagram. Here, in response to increasing hole-doping, the system retains the collinear AF order from the half-filling configuration, accommodating for the reduced filling level by the formation of charge-deficiency ``defects" in the lattice structure, i.e. localized lattice sites or series of adjacent lattice sites where the filling level is significantly lower than at the surrounding sites. This is distinguished from a charge-density wave (CDW) type of state in that the charge distribution is not continuously modulated and instead seems to be pinned to the underlying lattice (see $\langle n_i \rangle$ in Fig. \ref{fig: defect region}). The spin magnitude is directly modulated by the charge deficiencies, creating similar regions of lower spin magnitude corresponding to the regions of reduced charge, and there are some indications (see $\langle S_i \rangle$ in Fig. \ref{fig: defect region}) that these lines of spin/charge modulation serves as domain walls, separating $n_e=1.0$ Néel domains with differing Néel vectors. As the filling level is reduced further, the number of defect lines increase and their spacing on the lattice decrease, leading eventually to the semi-ordered collinear island domain phase (see Fig. \ref{fig: defect region}(b)). Here, we see the same type of abrupt charge depletion, but more ordered and seemingly pinned to the underlying lattice, respecting to some extent the four-fold rotational symmetry of the lattice. The site spins remain collinear.
The appearance of such spurious defects in the lattice, challenging to model by a regular restricted mean-field ansatz is a direct result of the unrestricted methodology used, and the formation of inhomogeneous configurations is a possible reason why previous research using translationally invariant ansätze have observed negative compressibility in this region of the phase diagram  \cite{Igoshev2010, Auerbach1991, Dongen1996, Stanescu2000}. 

The second main region of the phase diagram is characterized by well defined magnetic phases, i.e. magnetic and charge spatial modulation with clear and distinct ordering vectors. At half-filling, the system ground state is the well established Néel state. Away from half-filling for low to intermediate interaction strengths, magnetic phases of the type ($Q$, $\pi$) is prevalent, denoted by the blue region in Fig. \ref{fig: phase diagram alpha=0.0} (see in particular Fig. \ref{fig: phase diagram alpha=0.0} (c). Note that the phase is also associated with a stripe charge order and the pattern is in essence an incommensurate collinear spin-density wave (SDW). The presence of a ($Q$, $\pi$) in the low-interaction, doped Hubbard model has been reported by several sources \cite{Fresard1992, Igoshev2010} using homogeneous mean-field and slave-boson approaches, but without information on the charge distribution. This phase is often referred to as a ``spiral" magnetization, as can be intuitively be understood if one attributes the incommensurate \textit{Q}-component of the ordering vector solely to the relative orientation between spins and not magnitude modulation.

An interesting aspect of the differences in methodology in this paper compared to the restricted ansatz-type of mean-field analysis is while the same magnetic ordering vector $\boldsymbol{Q}$ can be predicted by both, for instance the ($Q$, $\pi$) phase, they indicate two very different states. Within a restricted mean-field methodology, one can choose for instance a spiral ansatz of the form $\langle \boldsymbol{S}_i \rangle = m[\cos (\boldsymbol{Q}\cdot\vecr_i), \sin(\boldsymbol{Q}\cdot\vecr_i),0]$  \cite{Igoshev2010, Kennedy2022}, omitting the \textit{z}-component for brevity. In such an analysis, \textit{m} and possibly $\boldsymbol{Q} $ are found self-consistently. Without the possibility for a spatially varying magnetization magnitude \textit{m}, a (\textit{Q}, $\pi$) phase is a spiral configuration, characterized by a spin-canted magnetization with a period determined by ${Q}$. Within our methodology however, the (\textit{Q}, $\pi$) phase can be a collinear SDW state with no spin-canting at all, two entirely different states. The incommensurate \textit{Q} component which causes spin-rotation in the restricted methodology instead represents an incommensurate charge modulation and subsequent spin-magnitude modulation in our system. As such, the comparison of ordering vectors originating within different methodologies should be done with caution, precisely due to the the additional possibility of having a varying magnetization magnitude in the present unrestricted framework.

An important feature in our diagram is the presence of higher-symmetry modifications of the same ordering vector. While the blue region is characterized by an ordering vector ($Q$, $\pi$), the green 2($Q$, $\pi$) region, ocurring at higher interaction strength, is characterized by the same type of ordering vector, but with twice the number of maxima in the magnetic structure factor, reflecting a state with higher symmetry. This is evident from considering the real-space magnetization pattern in Fig. \ref{fig: phase diagram alpha=0.0} (c) and (d).  

For $U\sim 5.0-7.0$, commensurate (0, $\pi$) and the incommensurate (0, $Q$) arise for filling levels $n_e\sim 0.3-0.5$. An interesting consequence of the unrestricted mean-field model is way in which the system transitions between two states. The violet region in-between the purple (0, $\pi$) and the blue ($Q$, $\pi$) region is simply a combination-phase where both ordering vectors to some extent are present in the ground state phase (see Fig. \ref{fig: phase diagram alpha=0.0} (e)). This is different from usual mean-field models where the \textit{a priori} selection of mean-fields cause abrupt transitions between phases as one type of phase becomes energetically unfavorable to another. 

\begin{figure*}[p]
\centering
\includegraphics[width=0.95\textwidth]{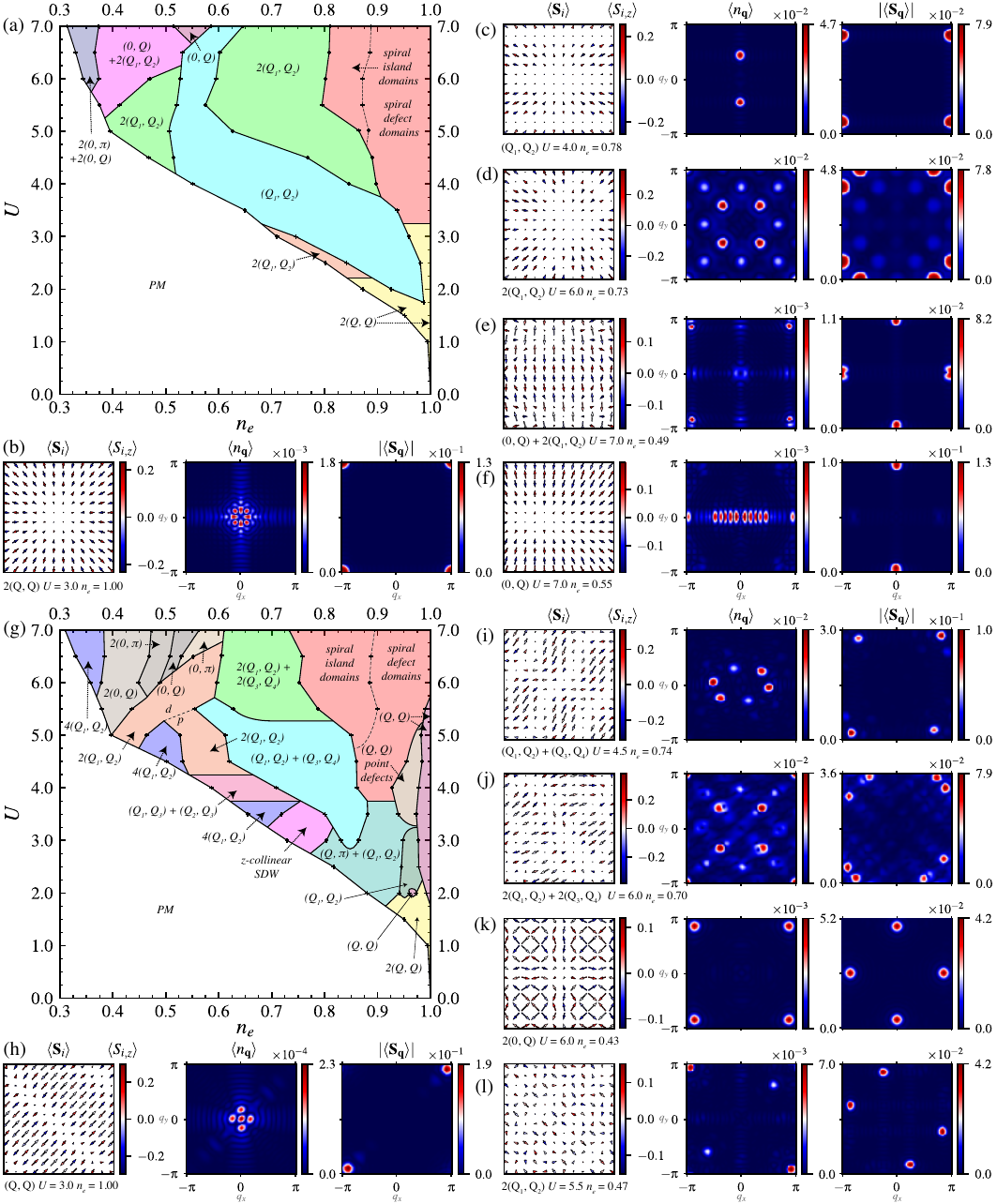}
\caption{Ground state magnetic phase diagrams obtained for a $24\times24$ lattice in with Rashba spin-orbit coupling $\alpha_R=0.10t$ (a) and $\alpha_R=0.25t$ (g) for the SSD mean field model. 
A selection of the ground state configurations are shown in the subplots (b)-(f) for the $\alpha_R=0.10t$ diagram and (h)-(l) for the $\alpha_R=0.25t$ diagram. Each phase is characterized by its real space spin distribution $\langle \vecS_i \rangle$, its charge modulation vector $\langle n_{\vecq}\rangle$ and spin structure factor $|\langle \vecS_{\vecq}\rangle|$. The \textit{d}/\textit{p} notation in the $\alpha_R=0.25$ is meant to distinguish between diagonal and parallel charge modulation within phases with the same magnetic structure factor.}
\label{fig: phase diagram alpha=0.25}
\end{figure*}

\subsubsection{Phase diagram in the presence of Rashba spin-orbit coupling – $\alpha_R=0.10$ / 0.25}
The phase diagram was calculated using the same parameter ranges as for the $\alpha_R = 0.00$ diagram, in the presence of Rashba SOC with strength $\alpha_R=0.10 $ and 0.25. We find that the introduction of the Rashba effect in the Hubbard model dramatically increases the complexity of the ground state behaviour, especially in the case of $\alpha_R=0.25$. The presence of spin-orbit coupling changes the characteristics of the phases already present in the $\alpha_R=0.00$ diagram, as well as introducing completely new phases. 

In the case of $\alpha_R=0.10$ (see Fig. \ref{fig: phase diagram alpha=0.25} (a)-(f)), the phase composition  
of the phase diagram, i.e. the presence of distinct phases in distinct regions, resembles that of the diagram without SOC. The main impact is that in the most prevalent phases of the diagram, the previously ($\alpha_R=0.00$) commensurate component $Q_{x,y}=\pi$ now has become incommensurate. This is for instance visible in the blue region ($Q_1$, $Q_2$) phase, corresponding to the $\alpha_R=0.00$ ($Q$, $\pi$) phase where the previously commensurate $\pi$-component now has moved slightly inwards from the 1st Briollouin zone (1BZ) boundary, effectively lifting the previous staggered order in the direction perpendicular to the charge modulation and introduced spin canting between adjacent spins. We also observe a considerable broadening of the magnetic structure factor maxima for this phase in particular, as compared to the more distinct and sharply defined ordering vectors in the absence of SOC. We argue that this might be due to the way SOC alters the shape of the Fermi surface, broadening the range of ordering vectors ($Q_1$, $Q_2$) at which nesting of the Fermi surface occurs. This is discussed in more detail in the following chapter on RPA and linear response. 

The previously discussed defect regions retains its main characteristics for $\alpha_R=0.10$. As before, we draw a distinction between the defect domains, characterized by spurious and randomly located charge deficiencies, and the island domains where the charge deficiencies align in a somewhat ordered manner, pinned to the underlying lattice structure. The main difference in this region upon the introduction of SOC is that the magnetization now becomes spiral, with spin-canting occurring between spins on adjacent sites, as opposed to the collinear order in the absence of SOC. The emergence of defect lines and features closely adhering to the underlying lattice is very similar to the behaviour in the absence of SOC, and given the magnitude of the Hubbard-\textit{U} compared to $\alpha_R$ in this region, we argue that this is inherently a property of the regular Hubbard model, being only slightly modified by the introduction of SOC.

At half-filling, the $\alpha_R=0.10$ system displays a diagonal 2($Q$, $Q$) order, which can be thought of as the regular Néel state, but with the diagonal ($\pi$, $\pi$) ordering moving in towards the center of the 1BZ, becoming incommensurate. This is in apparent agreement with the predictions of Kawano \etal\ \cite{Kawano2023} which predicts a ($Q$, $Q$) state at half-filling for systems with Rashba SOC strength comparable with ours, hinting towards the prevalence of a higher symmetry 2($Q$, $Q$) for systems with relatively low SOC strengths. Our finding of a 2($Q$, $Q$) state for $\alpha_R=0.10$ is therefore likely not in violation of their findings. We also observe the half-filling ground state to remain stable away from half-filling, remaining the system ground state at low interaction strengths as the system is doped. 

The doped region towards higher interaction strengths retains some of the characteristics of the diagram without SOC, the main difference being that the $2(Q_1, Q_2)$ state remains favourable (present in the $(0,\text{ }Q)+2(Q_1, \text{ }Q_2)$ combination phase) all the way to $U=7.0$ while the related $2(Q,\text{ }\pi)$ phase at $\alpha_R=0.0$ is replaced by stripe-like configurations at these interaction strengths. 

\begin{figure}
    \centering
    \includegraphics{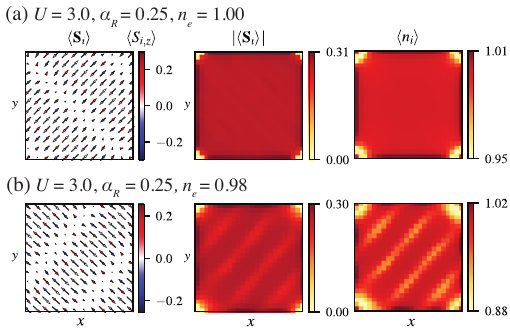}
    \caption{The ($Q$, $Q$) phase at half-filling for $U=3.0$, $\alpha_R=0.25$. Upon reducing the filling level, a SDW phase arises as the charge density $\langle n_i \rangle$ is spatially modulated with a subsequent modulation in the spin magnitude $|\langle \vecS_i \rangle |$.}
    \label{fig: half-filling}
\end{figure}

\begin{figure}[hbt]
    \centering
    \includegraphics{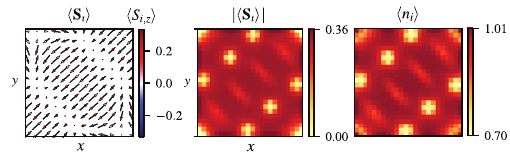}
    \caption{The $(Q,Q)$ point-defect phase at $n_e=0.97$, $U=4.0$ present in the $\alpha_R=0.25$ phase diagram (Fig. \ref{fig: phase diagram alpha=0.25}), emerging as the (\textit{Q}, \textit{Q}) phase at half-filling is doped for \textit{U} between 3.0 and 5.5. The key distinction from the domain-type phases at higher interaction strengths is the degree to which the charge depletion regions are ordered.}
    \label{fig: point defect}
\end{figure}

For $\alpha_R=0.25$, the phase composition of the model changes quite significantly with the introduction of several new phases. To begin, we observe the lower symmetry ($Q$, $Q$) state at half-filling for interaction strengths above $U=1.5$, replaced by the higher symmetry 2($Q$, $Q$) state from the $\alpha_R=0.10$ diagram below this value. As with for the $\alpha_R=0.10$ system, the half-filling configuration persists to some extent as ground state as the system is doped. The half-filling ground states do not therefore show the same instability upon doping that is characteristic for the Néel state. At half-filling, the ground state has a spatially constant spin magnitude, but in response to doping, a SDW state emerges (see Fig. \ref{fig: half-filling}). Note however that this phase is ultimately also susceptible to defect formation, a new emerging phase in the $\alpha_R=0.25$ diagram being the (\textit{Q}, \textit{Q}) point defect phase where the CDW charge ordering of the doped (\textit{Q}, \textit{Q}) phase breaks down, creating systematically ordered charge depletion spots in the lattice, see Fig. \ref{fig: point defect}. This phase is distinguished from the defect phases at higher interaction strength in that the charge defect formation to some extent follows the prevailing magnetic ordering in the vicinity of the phase. 

The central region of the ($Q_1$, $Q_2$) phase of the $\alpha_R = 0.10$ diagram becomes the more complicated ($Q_1$, $Q_2$) + ($Q_3$, $Q_4$) state at $\alpha_R=0.25$ (see Fig. \ref{fig: phase diagram alpha=0.25}(i)). Towards lower filling level, the ($Q_1$, $Q_2$) $\alpha_R=0.10$ shows an intricate dependence on \textit{U} and $n_e$ for $\alpha_R=0.25$, transforming into an array of different phases. At higher interaction strengths for lower filling levels $n_e= 0.3$--$0.6$, the phase composition of the diagram becomes even richer with a selection of commensurate and incommensurate stripe phases in both high- and low-symmetry variants ($2(0, \text{ }Q)$ vs. $(0, \text{ }Q)$ etc.). In this region, $U\gg\alpha_R$, and the fact that the increase in SOC strength introduces such a significant increase in phase composition points to the large near-degeneracy of the Hubbard model ground states, causing a slight perturbation in model parameters to cause a significant change in ground state behaviour. We note that a key feature of unrestricted mean-field techniques is that it allows the system to freely choose its configuration, which is a significant benefit of the method. However, the same freedom makes characterization of the phases much more challenging, as the possible phases themselves change significantly with the model parameters.

As far as we know, the only published phase diagrams for the doped Rashba-Hubbard model is by Kennedy\ \etal\ \cite{Kennedy2022} and Beyer \etal\ \cite{beyer_et}. Using a mean-field technique with a spiral ansatz, the methodology used by Kennedy \etal is not capable of assessing the direct impact of Rashba SOC on the magnetic phases (such as the change from ($Q$, $\pi$) to ($Q_1$, $Q_2$) phase when SOC is turned on), only the changes in energetic favourability between the different, pre-established phases. This might be a central reason why they, in conflict with our findings as well as Kawano \etal\ \cite{Kawano2023}, predict the half-filling Néel state to persist as SOC is introduced. Finding also the ground state phases for a fixed density \textit{n}, there is also the risk of thermodynamic instabilities as previous research using a similar methodology and ansatz has revealed, i.e. by observing negative electron compressibility close to half-filling \cite{Igoshev2010}. It is in particular near half-filling that we find the most challenging phases to characterize, such as the various defect configurations, alluding to the challenging system properties in this region. Beyer \etal\; predict a combination of commensurate and incommensurate SDW ground states in the Rashba-Hubbard model for filling levels between $n_e=0.45$ and 0.55 for SOC strengths relevant for this paper, but do not find any signs of charge-density wave (CDW) formation in their study, in contrast with our findings where intertwined CDW and SDW formation is predicted to have a significant presence in the model ground state, both in absence and presence of SOC.

\begin{figure}[t]
    \centering
    \includegraphics{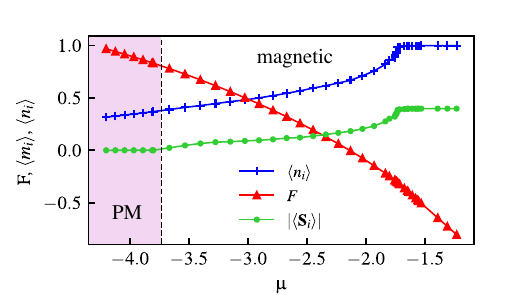}
    \caption{The free energy $F$, average filling level $\langle n_i \rangle$ and average magnetization $|\langle \boldsymbol{S}_i\rangle|$ for the $U=5.0$, $\alpha_R=0.25$ ground-state plotted as a function of the chemical potential $\mu$. Both the transitions between magnetic phases as well as the magnetic-paramagnetic transition is continuous. The black stippled line denotes the defined paramagnetic transition occurring when $|\langle \boldsymbol{S}_i\rangle|$ falls below 0.01.} 
    \label{fig: free energy} 
\end{figure}
\begin{figure}[h]
    \centering
    \includegraphics{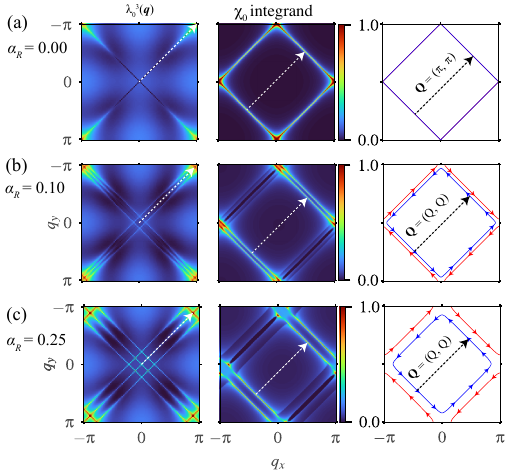}
    \caption{Largest eigenvalue $\lambda_0^3(\vecq)$ of the bare magnetic susceptibility, contribution of the respective nesting vector $\boldsymbol{Q}$ from the Fermi surface as well as the Fermi surface for the half-filled system ($\mu=0.0$), plotted for (a) $\alpha_R=0.00,\text{ (b) }0.10, \text{ and (c) } 0.25$. Note that the ordering vector at which the susceptibility diverges, ($\pi,\text{ }\pi$) in the absence of SOC, becomes incommensurate as $\boldsymbol{Q}=(Q, \text{ }Q)$ with the magnitude of \textit{Q} decreasing with increasing $\alpha_R$, corresponding to an increasing spatial period of the magnetic modulation. 
    }
    \label{fig: half-filling susceptibility}
\end{figure}

\begin{figure}
    \centering
    \includegraphics{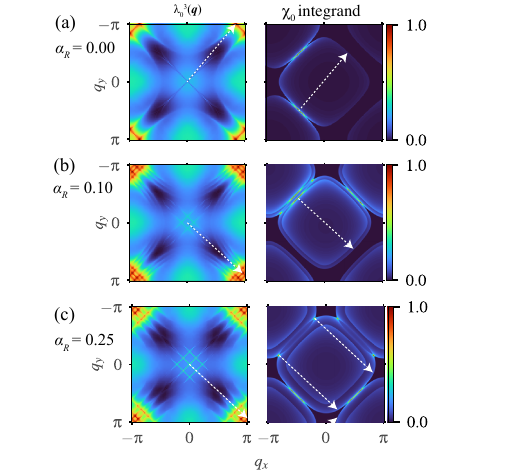}
    \caption{Largest eigenvalue $\lambda_0^3(\vecq)$ of the bare magnetic susceptibility, contribution of the respective nesting vector $\boldsymbol{Q}$ from the Fermi surface as well as the Fermi surface for the system with chemical potential $\mu=-0.5$, plotted for (a) $\alpha_R=0.00,\text{ (b) }0.10, \text{ and (c) } 0.25$. The Fermi surface with nesting vectors are here omitted due to the increased complexity of the ordering vectors in the doped model, making it challenging to pin-point the correct manner of the nesting.}
    \label{fig: doped susceptibility}
\end{figure}

\subsubsection{Phase transitions, fluctuations and free energy}

We briefly comment on the nature of the phase transitions in our system. The free energy, average system magnetization and filling level is shown for $U=5.0$, $\alpha=0.25$ in Fig. \ref{fig: free energy}. The stippled line denotes the magnetic-paramagnetic transitions, defined to occur when the average system magnetization drops below 0.01. As is evident, the transition from a magnetic configuration to the paramagnetic state is a continuous phase transition, denoted by the vanishing magnetization of the ground state. Note that the transition between the different magnetic configurations is also continuous. Given the second-order nature of these phase transitions, it is likely that fluctuations would affect phase boundaries \cite{Igoshev2010}, both between magnetic phases and for the magnetic-paramagnetic transition.

The phase diagrams shown in Fig. \ref{fig: phase diagram alpha=0.0} and \ref{fig: phase diagram alpha=0.25} show a diverse composition of phases. In particular, the $\alpha_R=0.25$ diagram is complex with many competing orders, which can partly be attributed to the previously discussed SOC-induced Fermi surface nesting. There is a well known tendency of mean-field type frameworks to overestimate ordering \cite{overestimate, overestimate_2}, which is important to keep in mind when assessing the phase diagram. As mentioned, the treatment of fluctuations is inherently absent within the mean-field formalism and the inclusion of fluctuations is expected to have an effect on the phase boundaries and possibly on the number of distinct phases observed in the diagram. There are limited phase diagrams published on the Rashba-Hubbard model which makes assessing the potential impact of fluctuations on the phase composition challenging. However, we note that in the half-filling limit, our approach closely reproduces the results reported by Kawano \etal~\cite{Kawano2023}. They used a density matrix embedding theory which more accurately accounts for electron correlations. This similarity suggests the validity of our framework in the half-filled limit.

\subsubsection{RPA and Fermi surface nesting}
Upon the introduction of Rashba spin-orbit coupling, the Fermi surface is altered significantly. Nesting of the Fermi surface (FS) is an important mechanism in establishing lattice superstructure such as spin- and charge-density waves. The mechanism behind the formation of magnetic phases in the weak-coupling limit can be understood by considering the magnetic susceptibility in the RPA framework. In Fig. \ref{fig: half-filling susceptibility}, the largest eigenvalue $\lambda_0^3(\vecq)$ of the bare susceptibility matrix is plotted as a function of magnetic ordering vector $\vecq$. 
The $\vecq$-vector causing the largest eigenvalue is then denoted the dominant magnetic ordering vector $\boldsymbol{Q}$. The contributions due to nesting of the non-interacting FS with nesting vector $\boldsymbol{Q}$ from different regions of the FS, causing the susceptibility to diverge, is then mapped by considering the contributions of the specific ordering vector to the susceptibility integrand in Eq. \ref{eqn: susceptibility integrand}, evaluated across the 1BZ. Finally, the non-interacting FS itself is plotted with the nesting vector $\boldsymbol{Q}$. In Fig. \ref{fig: half-filling susceptibility}, this is shown at half-filling for $\alpha_R = 0.00$, 0.10 and 0.25. From the figure, it is apparent that as the spin-orbit interaction is turned on and increased, $\boldsymbol{Q}$ transitions from the initial ($\pi$, $\pi$) state to a diagonal ($Q$, $Q$) state with $Q_{\alpha_R=0.25} < Q_{\alpha_R=0.10}$. For $\alpha_R \neq 0.0$, the spin-degeneracy of the non-interacting FS is lifted and we observe that the dominant nesting vector describes nesting between states in the same FS, but with opposite spin due to the spin-momentum locking property of spin-orbit coupling. 

Fig. \ref{fig: doped susceptibility} shows the same properties, but for a system doped to $\mu=-0.5$. We note that the chemical potential in the RPA framwork is different from the actual chemical potential in our SSD system due to the Hubbard-\textit{U}-induced shift in the chemical potential. The qualitative properties of the doped susceptibility is however applicable. We observe that in the absence of SOC, the dominant ordering vector becomes $\boldsymbol{Q}=(Q$, $\pi)$ where one component in effect has become incommensurate. This can intuitively be understood by considering how the reduction in filling level breaks the ``perfect" nesting of the half-filled FS, causing the nesting to become imperfect, occurring only in selected regions. As the system is doped further, the magnitude of the incommensurate component decreases. This is in accordance with the observed $Q$ behaviour which decreases from $Q\sim\pi$ towards 0 as the system is doped, giving rise to the stripe phases ($0$, $Q$) / (0, $\pi$) (see for instance Fig. \ref{fig: phase diagram alpha=0.0}). For $\alpha_R=0.10$, the second ordering vector component becomes incommensurate as well, but the ordering vector remains off-diagonal on the form ($Q_1$, $Q_2$) as opposed to the diagonal half-filling form ($Q$, $Q$). Finally, in the $\alpha_R=0.25$, the dominant ordering vector regains the ($Q$, $\pi$) form, but the susceptibility shows divergence also for several other ordering vectors. 

An important take-away from the RPA analysis is how the magnetic susceptibility and its $\vecq$-dependence becomes significantly more complex in the presence of SOC. The appearance of several unique and distinct ordering vectors, especially relevant for the doped susceptibility, explains to some extent the richness of the $\alpha_R\neq 0$ diagrams. We note that while the RPA analysis is generally valid for low interaction strengths, it can still give us a qualitative understanding of the driving mechanisms behind the emergence of magnetic order, both in the absence and presence of spin-orbit coupling. 

\subsection{Self-consistency calculation and convergence\label{sec: SF}}
For each combination of model parameters $U$, $\alpha_R$, and $\mu$, an initial 10 independent self-consistency calculations were performed with the spin-distribution randomized in each calculation. The converged phase with lowest free energy were chosen as the ground state configuration. The initial charge distribution was set to $\langle n_i \rangle=1$ across all lattice sites.

A measure of the stability of the calculations is the deviation in average filling level and magnetization across the 10 independent calculations for a  given set of parameters, serving as an indication of whether the same energy minimum is reached consistently independent of initial conditions. We generally observed a larger tendency towards ordered phases and high calculation stability for interaction strengths below $U=5.0$ and below $n_e=0.7$ for interaction strengths above $U=5.0$. In these more well-behaved regions, the self-consistency algorithm typically converged to ordered phases in good coherence with the ``trend" observed for similar model parameters. For $U>5.0$ and $n_e >0.7$, the converged phases showed a higher degree of unpredictability, both in terms of the magnetic ordering of the phase as well as the average density of the phase. The obtained ground state after 10 calculations was often not satisfactorily ordered, and with the use of an unrestricted ansatz, order is not a criterion for convergence. The system may become stuck in an energy landscape riddled with local minima, preventing the configuration from reaching the ``correct" ground state. This observed inability of the system to establish an ordered state is likely linked to the high interaction strength which, due to the unrestricted mean-field ansatz, prevents the randomized initial spin distribution from redistributing properly in order to establish an ordered state. In effect, we end up with semi-ordered states with lattice defects as remnants from the randomized initial distribution. 
\begin{figure}[htb]
    \centering
    \includegraphics{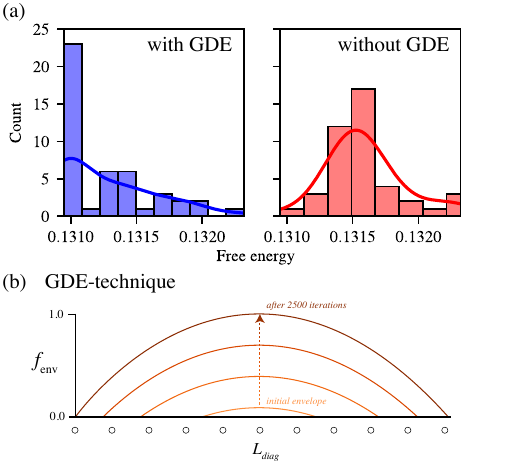}
    \caption{(a) The distribution of ground states, characterized by their free energy, after 45 independent self-consistency calculations with and without the gradual deformed envelope (GDE) technique for the model parameters $U=6.0$, $\alpha_R=0.00$, $\mu=-2.2919$. The continuous line denotes a distribution estimation based on the kernel density estimation (KDE) method. Half of the simulations with GDE achieve an approximate ground state with free energy $F \leq 0.13105$ and only 2.2 \% of the calculations without GDE fell below this 50 \% percentile. We argue that this limited study indicates  a generally observed trend which is that calculations with GDE consistently obtain lower energy configurations. (b) A schematic of the GDE technique showing the magnitude of the envelop $f_\text{env}$ across the lattice diagonal $L_{diag}$. In the first 2500 iterations, the envelope is linearly increased towards the final shape discussed in Sec. \ref{sec: SSD}. Initially, the envelope is non-zero only at 4-8 sites at the lattice center, allowing the magnetization initially established there to act as a seed for the rest of the lattice.}
    \label{fig: gde}
\end{figure}

In response to this, we built on the SSD technique and developed a gradual deformed envelope (GDE) technique. In essence, with GDE we change the height of the envelope throughout the self-consistency calculations in order to ``grow" the magnetic configuration from the centre point of the lattice in a controlled manner, thus avoiding the situation where the magnetization in two separate regions of the lattice develop independently and inconsistently, causing lattice defects where they eventually intersect. The motivation for this technique is taken from how real-world liquids crystallizes on seed-crystals upon solidification in order to form a coherent monocrystalline material. When using the GDE technique, at the onset of the calculation, the envelope is only non-zero at the center of the lattice, causing the magnetization to establish only in a tiny region consisting of 4-8 sites. The calculations are initiated with randomized initial magnetization as for the regular calculations. As the iterations progress, the envelope is effectively raised, increasing in magnitude and becoming non-zero in a continuously growing circle which eventually compasses the entire lattice. In this way, the spin distribution is given time to establish in the center before it steadily grows outwards toward the lattice edges analogous to how a monocrystalline material would solidify on a seeding crystal. The envelope was raised with a constant rate in the first 2500 iterations of the self-consistency calculations. After 2500 steps, the envelope was static and equivalent in shape to the envelope used in calculations without the GDE technique. Upon introduction of the technique, we observed a significant increase in the ability of the system to access lower-energy ground states. As an indicator of the effectiveness of GDE, we chose a particular combination of model parameters ($U=6.0$, $\alpha_R=0.00$, $\mu=-2.2919$, $n_e\simeq0.7$) in the high-interaction region troubled by poorly ordered solutions. We performed 45 independent calculations of the model ground state with the initial magnetization randomized between each calculation. The 45 calculations was performed both with and without the GDE technique, and the distribution of obtained ground states, characterized by the free energy of the configuration, is shown in Fig. \ref{fig: gde}. Using the GDE method, we find that half of the simulations manage to reach an approximate ground state with free energy $F \leq 0.13105$. For comparison, only $2\text{ }\%$ of the calculations without the GDE method converged to the same energy range. Thus, we conclude that the number of numerical experiments required to confidently identify the ground state energy in the high-interaction limit could possibly be reduced by well over an order of magnitude using our GDE approach. 

Finally, we employed an annealing technique in order to improve ordering. As discussed above, while the phases are well converged, in using an unrestricted ansatz, we have no guarantee that these phases are well ordered with long-range ordering across the lattice. As the site magnetization and charge number is determined self-consistently within the local environment of adjacent sites only, we can get an intuition about why it is so difficult to establish well-ordered phases with a coherent ordering across the entire lattice. Many of the ground state configurations showed a significant tendency towards a specific ordering vector, even if the order was not perfectly established. In order to overcome potential energy barriers , an annealing technique was used. In effect, the converged solutions were reinserted into the iterative algorithm, but with a higher initial system temperature $T$. This temperature was chosen as a small factor proportional to the interaction strength (typically $U \sim 0.01$--0.1). As the iterations proceeded, this temperature was linearly reduced to the original $T=0.01$, the idea being that the increased energy to the system would allow the lattice to reorganize and redistribute charge and magnetization before the temperature would be lowered again. This method increased the degree of ordering significantly, most significantly in the above mentioned regions with $U>5.0$, $n_e>0.7$.

\subsubsection{Finite size scaling and the effect of SSD}
From the outset of this paper, we discussed how we expected phases with incommensurate order to be prominent in the phase diagram of the Rashba-Hubbard model given their appearance in the ground state of the regular Hubbard model \cite{Schulz1990, Hiroyuki2016,Halboth2000, Kawano2023}. Assuming a magnetization mechanism driven by nesting common for mean-field systems, the introduction of SOC breaks the perfect ($\pi$, $\pi$) nesting giving rise to incommensurate ordering vectors. In addition to the challenge of obtaining the ``correct" model ground state, there is also always necessary to consider the size of the finite size system and its affect on the system properties. A larger system size is expected to more closely emulate the thermodynamic limit at the cost of being computationally more expensive than a smaller system. As such, a trade-off between system size and feasible computational cost has to be made in obtaining phase diagrams as presented in this paper.

We argue that the introduction of the SSD technique is an appropriate approach to overcome these challenges. As has already been discussed in detail, the SSD approach avoids bias with respect to real space periodicity of the magnetic configuration in addition to screening out boundary effects. We however also argue, in agreement with Refs. \cite{Kawano2023, Kawano2022, Hotta2013} that the introduction of SSD allows the system behaviour in the thermodynamic limit to be reached for smaller system sizes, thus lowering the computational cost necessary to emulate bulk-like conditions.

\begin{figure}
    \centering
    \includegraphics{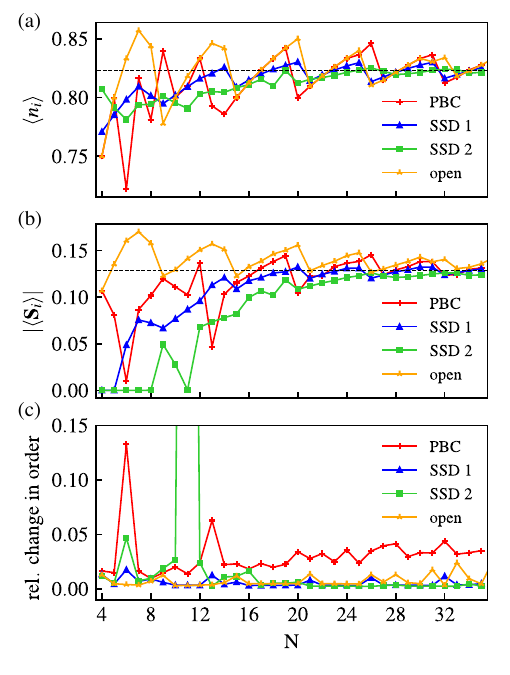}
    \caption{Finite-size scaling analysis for the ($Q_1$, $Q_2$) phase at $U=3.0$, $n_e\sim0.825$ in Fig. \ref{fig: phase diagram alpha=0.25} (a). The average system magnetization (a), filling level (b) as well as change in magnetic configuration quantified by the rel. change in order defined in Eq. (31) is shown for periodic boundary conditions (PBC), open boundary conditions as well as two types of SSD defined in the text. The stippled line is average of all data points above $N=20$ and is meant to represent a pseudo-thermodynamic limit. The results are based on 10 randomized trials for each data point and show that the SSD-type techniques converge more consistently towards the thermodynamic limit with less oscillatory behaviour compared to PBC and open boundaries.  In addition, both SSD and open systems show a significantly lower sensitivity in their magnetic configurations towards increasing system size compared with PBC. The outlier at $N=11$ for the SSD 2 system (c) takes the value 1.3191 and is attributed to a change in system ground-state, noted also by the collapse in magnetization (see $N=11$ in (b)).}
    \label{fig: finite size}
\end{figure}

In Fig. \ref{fig: finite size}, the ($Q_1$, $Q_2$) ground state configuration for $\alpha_R=0.10$, $U=3.0$, $n_e\sim 0.825$ [cf. Fig. \ref{fig: phase diagram alpha=0.25}(a)] was obtained for a system with (1) periodic boundary conditions (PBC), (2) open boundary conditions, and two variants of SSD. The data points are the free energy minimum of 10 randomized trials. SSD~1 corresponds to the technique employed in this paper where the envelope is zero at the lattice corners and a small finite envelope value remains at the lattice edges owing to the shape of the envelope. SSD~2 corresponds to a ``perfect" envelope where the envelope value is zero both at the lattice corners and edges. Due to the sinusoidal shape of the SSD envelope, this entails that a significant portion of the lattice sites close to corners are lost due to the circular contour of the envelope and thus that the effective system size is smaller than the actual $N\times N$ sites. In Fig. \ref{fig: finite size} (a), the average system filling level is shown, in (b) the average system magnetization and in (c) a relative change in order for lattices ranging from $4\times4$ to $35\times35$ in size. The relative change in order is defined by
\begin{equation}
     \text{rel. change in order }= \frac{\sum_{\boldsymbol{q}}\big||\boldsymbol{S}_{\boldsymbol{q}}|^{N}-|\boldsymbol{S}_{\boldsymbol{q}}|^{N-1}\big|}{\sum_{\boldsymbol{q}}|\boldsymbol{S}_{\boldsymbol{q}}|^{N-1}}
\end{equation}
and is a measure of the relative change in magnetic configuration as the system size is increased from $(N-1)\times(N-1)$ to $N \times N$. In this error estimate, the four-fold rotational symmetry of the square lattice is taken into consideration in order to account for identical configurations related by a rotation. It is evident from the average system magnetization and filing level as function of system size $N$ [Fig. \ref{fig: finite size} (a) and (b)] that the SSD-type techniques to a less extent experience periodic oscillations with increasing system size, something that characterizes both PBC and open systems. In addition, the lowest panel in Fig. \ref{fig: finite size} shows how the magnetic configuration with PBC in particular is sensitive to system size. The magnetic configuration of the open system showed low sensitivity to increasing system size, apart from the oscillating average magnetization and filling level of the phase, and behaved very similarly to SSD. This is likely because open and SSD type boundary conditions are quite similar in the sense that no matching of the magnetic pattern is required at the system edges. While the open system likely experiences some frustration at the lattice edges, it is still free to establish the desired periodicity in the lattice interior, setting it apart from PBC with the requirement of lattice matching at the edges likely imposes a stricter limitation on the realized configuration. 

Based on the above discussion which is deemed representative for the system behaviour as a whole, a lattice size of $24\times 24$ with SSD (see SSD~1 in Fig. \ref{fig: finite size}) is likely to emulate the thermodynamic limit, showing both less oscillatory behaviour in the magnetisation and filling level compared to PBC and open systems, in addition to a low rel. change in order for increasing system sizes. This indicates that the system converges to the ``correct'' ground state already for system sizes $N$ in the range 16--20, with filling levels and magnetization profiles that are considered representative of the thermodynamic limit. 

\subsection{Dynamical magnetic properties: quantum quenching}
\subsubsection{Interaction quench at half-filling}
Time dynamics of the magnetic configurations obtained in the self-consistent SSD framework was simulated solving the equations of motion defined by Eq. (\ref{eqn: number equation})-(\ref{eqn: spin-flip equation}). The relevant equation of motion for each of the correlations in the statistical initial state was solved numerically using a Runge Kutta method of order 4. A fixed timestep of $h=0.01$ was used for all calculations and the timesteps are in units of $\hbar/t$.

Starting with the equilibrium phase at half-filling, $U=6.0$ for $\alpha_R=0.00$ and 0.10 respectively, a quench towards lower interaction strengths was simulated using a timestep of $h=0.01$ and a Runge-Kutta method of order 4. As discussed in the section on transient dynamics in the SSD-model, the envelope-modulated system can in some sense be considered an open system, connected to a particle reservoir of limited particle number. When a self-consistent solution is obtained in the static calculations and an initial statistical state is generated to serve as the initial configuration for the quench, the total particle number is fixed. This entails that at the start of the temporal evolution, there exist a given number of electrons on the lattice, distributed between the interior bulk region and the reservoir-like edges. The consequence of this distinction between bulk and edge states is that if a quench changes the optimal bulk filling level, the system may to some extent alleviate this by moving electrons in and out of the interior region of the lattice, a transfer of electrons which would have been impossible in a regular, unmodulated closed system. In that sense, the particle number on the lattice is conserved, but the electrons which determine system observables,  i.e. the electrons in the interior region, is not. In order to simulate a quench of a half-filled system which is effectively closed and not able to exchange electrons with an exterior reservoir, we quench the system not only in \textit{U}, but also with the appropriate chemical potential of the half-filled state we quench towards. In effect, for a given $U$, we identify the corresponding $\mu$ giving half-filling and we quench towards this ($U$, $\mu$) pair. In this way, we cancel the \textit{U}-induced renormalization of the chemical potential, thus remaining at an approximately constant filling level. 
\begin{figure}
    \centering
    \includegraphics{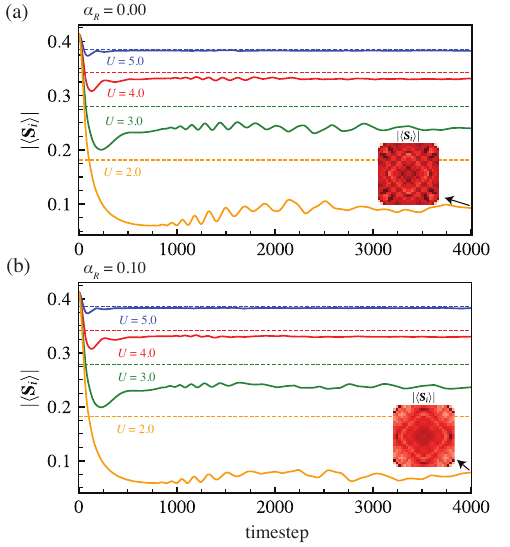}
    \caption{Magnetization magnitude dynamics following an interaction quench towards lower interaction strengths, starting from the $U=6.0$, half-filled configuration for (a) $\alpha_R=0.00$ and (b) $\alpha_R=0.10$. The inset shows the spatial magnetization magnitude distribution which shows signs of a quench-induced inhomogeneity not present in the equilibrium phases.}
    \label{fig: half-filling dynamics}
\end{figure}

\begin{figure}
    \centering
    \includegraphics{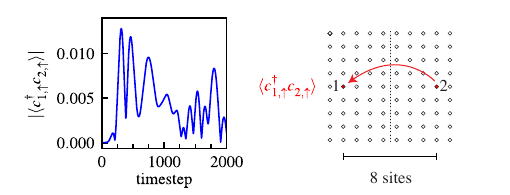}
    \caption{Amplitude of the expectation value of the non-local correlation $\langle\cc{1,\uparrow}\ca{2, \uparrow}\rangle$ for the arbitrarily labelled sites 1, 2 separated by 8 sites on the lattice across the lattice center. This specific correlation is from the quench towards $U=2.0$ in absence of SOC in Fig. \ref{fig: half-filling dynamics}. The non-local expectation value is of order $10^{-5}$ in the self-consistent equilibrium state, but quickly increases to $\sim10^{-2}$ as the system is evolved in time.}
    \label{fig: correlation}
\end{figure}
The effects of the interaction quench on the magnetization magnitude of the $\alpha_R=0.00$ and 0.10 systems are shown in Fig. \ref{fig: half-filling dynamics}.
An immediate observation is that the level at which the system magnetization stabilizes or oscillates around after the initial quench effects is significantly lower than the equilibrium magnetization, denoted by dotted lines. The equilibrium magnetization is the magnetization of the self-consistent phase obtained with the static framework with the same $U$, $\alpha_R$ and $\mu$. We also observe that while the initial magnetization response is of a coherent nature, i.e. taking on a damped sinusoidal shape, this is modified by the introduction of more high-frequent oscillations, showing up around timestep 1000. 
The frequency of the initial damped sinusoidal oscillation is the largest for the quench towards $U=5.0$, decreasing in frequency when the post-quench $U$ becomes lowers. The inset plots in Fig. \ref{fig: half-filling dynamics} show the spatial distribution of the magnetization magnitude $|\langle \boldsymbol{S}_i \rangle |$ for the quench towards $U=2.0$. The spatial distribution is non-homogeneous, but ordered, in contrast to the equilibrium phases at $U=2.0$, $n_e=1.0$, $\alpha_R \in \{ 0.00, 0.10 \}$ which have a homogeneous magnetization magnitude distribution.

Tsuji \etal\ \cite{Tsuji} predicted, using a half-filled Hubbard model and non-equilibrium dynamical mean-field theory, that upon an interaction quench towards lower $U$, the system magnetization becomes trapped at a non-equilibrium level above the thermal magnetization. They also predicted the discrepancy between the non-equilibrium magnetization and the thermal value to increase with the quench magnitude. The dynamics following the quench in our model displays this latter quality with the difference in magnetization increasing with quench magnitude. However, apart from this, the system magnetization shows a behaviour opposite to the one discussed in the previously mentioned work \cite{Tsuji}. After the interaction quench, the system magnetization magnitude levels out at a level below the one predicted at equilibrium. We argue that this significant deviation from previously reported results is a possible consequence of the quench being performed in an 
unrestricted framework.
The quench-induced non-homogeneous magnetization magnitude (insets in Fig. \ref{fig: half-filling dynamics}) could indicate that in response to the interaction quench, a non-homogeneous order is established as a means of relaxing frustrations caused by the abrupt change in system environment. We observe the non-homogeneous order to be strongest for the quench towards $U=2.0$, but this behaviour is universal across all quenches at half-filling. The quench-induced inhomogeneous order may be related to the build-up of non-local correlations on the lattice. Correlations over many sites (8 sites shown in Fig. \ref{fig: correlation}) are generally many orders of magnitude below nearest-neighbour- and on-site correlations at the onset of the temporal evolution. These non-local correlations increase many orders of magnitude as the system is evolved. The property that these non-local correlators are initially very low, but increase by many orders of magnitude as the system evolved, may explain why the dynamics evolves ``smoothly" for the first 500 time-steps before high-frequency components become prominent. In effect, given the low values of non-local correlations, it is reasonable to conclude that the initial magnetization correction following the quench is local, with each site adjusting only in response to its immediate neighbours. As time progresses, the whole lattice becomes more correlated through the dramatic increase in non-local correlators, possibly explaining why the system response becomes more complex and why a non-homogeneous magnetization order establishes. 
\begin{figure}
    \centering
    \includegraphics{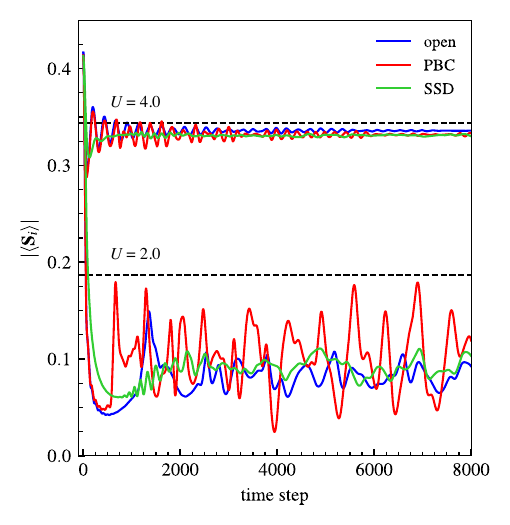}
    \caption{Magnetization magnitude dynamics of the $U=4.0$ and $U=2.0$ quenches from Fig. \ref{fig: half-filling dynamics} (a), performed (1) without SSD, but with open boundary conditions and (2) without SSD, but with periodic boundary conditions. The magnetization of the equilibrium phases is represented by a stippled line. We observe the same tendency towards undershooting the equilibrium magnetization value, while noting the the system response with periodic boundary conditions is significantly more oscillatory than for the system with open boundary conditions, with and without SSD. The results indicate that the quench-induced sub-equilibrium magnetization observed likely is an inherent property of the unrestricted mean-field ansatz employed.}
    \label{fig: pbc vs ssd vs open dynamics}
\end{figure}
\subsubsection{Quench at half-filling with (1) periodic boundary conditions and (2) open boundary conditions without SSD}
To further assess the cause of the breakdown in system magnetization and the possible impact of SSD on the dynamics of the system, the quench towards $U=2.0$ and $U=4.0$ from Fig. \ref{fig: half-filling dynamics} was performed also (1) without SSD, but with regular open boundary conditions, and (2) without SSD, but with periodic boundary conditions (PBC). For both (1) and (2), the system was prepared with the same conditions as the subsequent quench, i.e. the PBC quench was simulated using an initial state obtained self-consistently using PBC and so on. The comparison between the quenches using the three different conditions are shown in Fig. \ref{fig: pbc vs ssd vs open dynamics} and we argue that the results indicate that the observed behaviour in Fig. \ref{fig: half-filling dynamics} is not a result of the SSD methodology, but rather a result of the unrestricted mean-field ansatz employed in this paper, in line with the discussion above on the emergence of inhomogeneity. Tsuji \etal\ employ a non-equilibrium dynamical mean-field model which, while it treats spatial correlations in a form similar to regular Hartree Fock mean-field theory, also treats temporal correlations \cite{noneq_dynamics}. This method is more advanced than the regular mean-field technique used in this paper, but there has been previously published literature on the use of mean-field theory together with the Heisenberg equations for simulating quantum quenches, for instance on gap dynamics in the Letter by Peronaci \etal~\cite{Peronaci}. It is thus not obvious that the mean-field methodology is inadequate for this type of system dynamics. 

A significant difference between Tsuji \etal, Peronaci \etal\ and this paper is the use of an unrestricted mean-field ansatz in the present case. It is plausible that it is the high number of degrees of freedom attributed to the site-dependent charge and magnetization which is responsible for the more turbulent response of our system to quenches compared to the above mentioned papers. The average magnetization shown in Figs. 12, 14 and 15 are an average over $24\times24$ sites and as such, it is reasonable to expect the system response to be complex in response to an abrupt change in environment conditions. Real-world systems typically involve an immense number of degrees of freedom and it is not unreasonable to expect that the response of such a system to an abrupt change in environment conditions also involve a complex relationship, owing to the intricate coordination of system degrees of freedom, and not necessarily a coherent response, typically observed in dynamics simulations using a restricted $\boldsymbol{k}$-space ansatz. 
\begin{figure}
    \centering
    \includegraphics{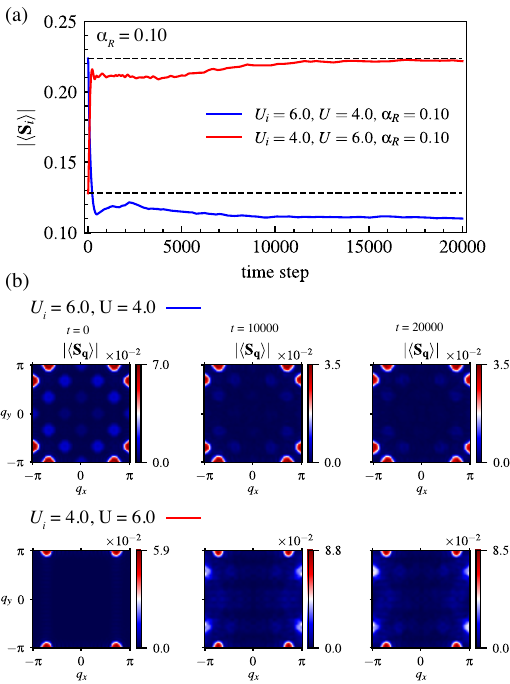}
    \caption{Magnetization dynamics following a quench between the 2($Q$, $\pi$) phase at $U=6.0$ and the ($Q$, $\pi$) phase at $U=4.0$ at $n_e \sim0.7$, $\alpha_R=0.10$. (a) The dotted lines denote the equilibrium magnetization of the two phases and we observe that the quench from $U_i=4.0$ initially undershoots the equilibrium magnetization of the $U=6.0$ phase before slowly approaching it, while the quench in the opposite direction does not approach the equilibrium magnetization within the number of timesteps. (b) The magnetic and charge structure factors are shown for $t=0$, 10000 and 20000 showing that while the system responds quickly in terms of adjusting magnetization magnitude, the magnetization pattern responds very slowly in comparison. 
    }
    \label{fig: doped quench}
\end{figure}
\subsubsection{Interaction quench of the doped ($Q$, $\pi$) / 2(\textit{Q}, $\pi$) phases}
A quench in the doped $\alpha_R=0.10$ system was performed between the ($Q_1$, $Q_2$) at $U=4.0$ and the higher-symmetry 2($Q_1$, $Q_2$) phase at $U=6.0$ at filling level $n_e\sim0.7$, effectively crossing the phase boundary between the two phases in Fig. \ref{fig: phase diagram alpha=0.25}(a) dynamically. The magnetization magnitude and magnetic structure factors following the quench are shown in Fig. \ref{fig: doped quench}. We observe the same tendency of the non-equilibrium magnetization falling significantly below the equilibrium magnetization while in the quench towards $U=6.0$, the magnetization approaches its equilibrium value eventually. This points to an important aspect of dynamics in the unrestricted model. Namely,  with the large number of independent degrees of freedom at play, the system response is complex in the sense that different system properties adjust on potentially very different timescales. While the magnetization responds quickly to the quench, this initial response phase characterized by a dramatic change in the magnitude, is replaced by a slow, drift-like evolution where frustrations are revealed and where the spins on individual sites adjust to adjacent lattice sites and to the rest of the lattice through the build-up of non-local correlations which also here is prominent. 

A key observation in the quench between the 2($Q_1$, $Q_2$) and ($Q_1$, $Q_2$) phases is that while the phase magnetization magnitude adjusts quickly, the ordering vector characterizing the phases does not and the phases retains their ordering relatively long after the quench is performed. A distinction is to be made between the quench from high-symmetry 2($Q_1$, $Q_2$) to low-symmetry ($Q_1$, $Q_2$) (blue line in Fig. \ref{fig: doped quench}), and the opposite (red line in Fig. \ref{fig: doped quench}).  The 2($Q_1$, $Q_2$) phase retains its ordering when quenched to $U=4.0$, showing instead a significant deviation in magnetization compared to the equilibrium phase. The opposite quench from ($Q_1$, $Q_2$) likewise retains its ordering initially, but as evident from the structure factors in Fig. \ref{fig: doped quench}, the higher symmetry phase slowly emerges as the system is evolved. It is possible that the emergence of the 2($Q_1$, $Q_2$) ordering coincides with the non-equilibrium magnetization approaching the equilibrium value and that the deviation between non-equilibrium and equilibrium magnetization is affected by the quenched phases retaining their original ordering. We can not rule out that in the quench from high to low symmetry 2($Q_1$, $Q_2$)$\rightarrow$($Q_1$, $Q_2$), the lower-symmetric ($Q_1$, $Q_2$) phase will emerge eventually. It is however evident that there is a asymmetry in the time-scale between the two opposite processes.

\section{Conclusions}
In summary, we have studied the ground state properties of the Rashba-Hubbard model on a square lattice with nearest neighbour hopping using an unrestricted mean-field charge and spin ansatz within a sine-square deformed envelope framework. We have shown that the introduction of Rashba spin-orbit coupling dramatically alters the phase composition in the model ground state, both through the modification of existing phases and by introducing completely new phases not present in the regular Hubbard model. Large parts of the phase diagrams are characterized by a rich combination of spin and charge order, verifying the need for a method which can characterize both. We have laid out in detail suitable methods increasing the ability to reach ordered and plausible ground states in the self-consistency calculation and have introduced the gradual deformed envelope (GDE) technique. In addition to the equilibrium study, we establish a framework based on the Heisenberg equation of motion for the study of magnetization dynamics in the model following instantaneous quenches in model parameters. We find that interaction quenches in the half-filled model induces a inhomogeneous spin-magnitude not present in the equilibrium phases. In addition, we observe a metastable system magnetization magnitude well below the magnetization predicted by the ground state phase diagram, possibly related to the build-up of non-local correlations on the lattice and the induced spin inhomogeneity. We also observe an asymmetry in timescales when quenching between a high and low symmetry phase in the doped system, finding the emergence of the higher symmetric state to occur at a timescale significantly shorter than for the opposite process.

\begin{acknowledgments}
We acknowledge funding via the Research Council of Norway Grant numbers 323766, as well as through its Centres of Excellence funding scheme, project number 262633. The numerical calculations were performed on resources provided by 
Sigma2, Project No. NN9577K - the National Infrastructure for High Performance Computing and Data Storage in Norway
\end{acknowledgments}

\bibliography{refs.bib}

\end{document}